\documentclass[amssymb,pra,twocolumn,longbibliography,notitlepage,superscriptaddress]{revtex4-1}

\newcommand{\kets}[1]{| #1 \rangle}
\newcommand{\bras}[1]{\langle #1 |}

\newcommand{\bt}{\boldsymbol{\theta}}
\newcommand{\bc}{\boldsymbol{c}}

\usepackage{float}
\usepackage[%
colorlinks=true,
urlcolor=blue,
linkcolor=blue,
citecolor=blue
]{hyperref}

\usepackage{color}

\usepackage{amsmath}                          
\usepackage{amsfonts}                         
\usepackage{amssymb}                          
\usepackage{amsthm}                           
\usepackage{mathdots}                         

\usepackage[braket, qm]{qcircuit}   

\usepackage{graphicx}                         
\usepackage{subfigure}                        
\usepackage{overpic}

\begin{document}

\title{Variational quantum eigensolvers by variance minimization}
\author{Dan-Bo Zhang}
\email{dbzhang@m.scnu.edu.cn}
\affiliation{Guangdong Provincial Key Laboratory of Quantum Engineering and Quantum Materials, GPETR Center for Quantum Precision Measurement and SPTE, South China Normal University, Guangzhou 510006, China}
\affiliation{Frontier Research Institute for Physics,
	South China Normal University, Guangzhou 510006, China}
\author{Zhan-Hao Yuan}
\affiliation{School of Physics, Nanjing University, Nanjing 210093, China}
\author{Tao Yin}
\email{tao.yin@artiste-qb.net}
\affiliation{Yuntao Quantum Technologies, Shenzhen, 518000, China}

\begin{abstract}
Variational quantum eigensolver~(VQE) typically minimizes energy with hybrid quantum-classical optimization, which aims to find the ground state.
Here, we propose a VQE by minimizing energy variance, which is called as variance-VQE~(VVQE). The VVQE can be viewed as an self-verifying eigensolver for arbitrary eigenstate by designing, since an eigenstate for a Hamiltonian should have zero energy variance. We demonstrate properties and advantages of VVQE for solving a set of excited states with quantum chemistry problems. Remarkably, we show that optimization of a combination of energy and variance may be more efficient to find low-energy excited states than those of minimizing energy or variance alone. We further reveal that the optimization can be boosted with stochastic gradient descent by Hamiltonian sampling, which uses only a few terms of the Hamiltonian and thus significantly reduces the quantum resource for evaluating variance and its gradients.

\end{abstract}
\maketitle
\section{Introduction}
The variational principle is instrumental for understanding physical theories and also becomes a powerful tool for solving computational physics problems. In recent years, the power of quantum computing of near-term noisy quantum processors is expected to be exploited with variational methods~\cite{yung_14,farhi_14,mcclean_16,o’malley_16,li_17,mcclean_17,shen_17,kandala_17,hempel_18,anschuetz_18,mitarai_18,moll_18,endo2018variational,kokail_19,takeshita_19,mcardle_19,Higgott_19,wu_19,verdon_19,liu_19,liu_19_gibbs,benedetti_19,grimsley_19,chowdhury2020variational,wang_20,danbo_20}, which refers to hybrid quantum-classical algorithms for optimization. A representative example is variational quantum eigensolver~(VQE)~\cite{yung_14,mcclean_16,o’malley_16,kandala_17,mcardle_19,liu_19,grimsley_19}, aiming to solve eigenstates for a quantum system that receives special interests due to its fundamental roles in quantum chemistry~\cite{yung_14,mcclean_16,o’malley_16,kandala_17,grimsley_19,arute2020hartreefock}, quantum many-body systems~\cite{liu_19,kokail_19,dallairedemers2020application}, and many other applications~\cite{anschuetz_18,xu2019variational,lubasch_20}.

Interestingly, while called as eigensolver, VQE typically solve the ground state of a Hamiltonian, using the energy as a cost function to optimize. From  the variational principle, such a cost function can gives an upper bound to the ground state energy.
For being an eigensolver, many variants of VQE have been developed to solve excited states, which can be understood as minimizing energy in a subspace, constrained by symmetries~\cite{greenediniz2019generalized}, enforced by orthogonality~\cite{nakanishi_19}, or obtained by eliminating the space of lower energy states~\cite{Higgott_19}, or using quantum subspace expansion~\cite{mcclean_17}.

While those variants of VQE can be successful to some extent, an alternative approach is to directly access whether a state is an eigenstate and use this criteria as a cost function to optimize. A simple answer for this is to use zero energy variance to justify an eigenstate since an eigenstate should have zero energy variance. Variational methods based on the zero-energy variance principle~\cite{bartlett_35,siringo_05,umrigar_88,umrigar_05,khemani_16,pollmann_16,vicentini_19}, in fact, can be originated to the early age of quantum mechanics for solving the Helium atom~\cite{bartlett_35}. Remarkably, zero variance puts a very strong constraint of the wavefunction structure, and it can be powerful for solving quantum many-body problems when well-approximated wavefunction is required~\cite{pollmann_16}. The variance principal has been applied experimentally to self-verify if a quantum simulator can prepare a ground state for a Hamiltonian~\cite{kokail_19}. It is, however,
awaiting for exploiting the power of quantum computing to develop a variational quantum eigensolver that is based on minimizing energy variance.

In this paper, we develop a variational quantum eigensolver based on minimizing energy variance, which we call as variance-VQE. For solving excited states of a Hamiltonian, we give two approaches, which represent eigenstates in different ways. One uses a single wavefunction ansatz, where different parameters correspond to different eigenstates.  The other uses a set of orthogonal wavefunction ansatz, and incorporates energy variances into one cost function to optimize.  We  numerically solve the energy potential surface for molecules, which is a fundamental quantum chemistry problem, and demonstrate properties and advantages of variance VQE.
We also show that optimizing a combination of energy and variance may be more efficient in finding a set of eigenstates with lowest energies, compared with one that optimizes energy or variance alone.
Moreover, we also investigate stochastic gradient descent for optimizing VVQE with Hamiltonian sampling, which can be useful for optimization with less quantum resources. Our work demonstrates that VQE by minimizing energy variance can be useful for calculating excited states, and also work out an avenue for efficient optimization.

The paper is organized as follows. We first propose variance VQE in Sec.~\ref{sec:vvqe}, and then apply it to solve excited states for molecules in Sec.~\ref{sec:excited_states}. In Sec.~\ref{sec:hamiltonian sampling}, we investigate the stochastic gradient for optimizing VVQE. Finally, we give a summary in Sec.~\ref{sec:summary}.

\section{Variance VQE}\label{sec:vvqe}
In this section, we first review the traditional approach of variational quantum eigensolver that minimizes the energy. Then, we formulate a new type of VQE that minimizes the energy variance and then discuss its optimization.

\subsection{VQE by energy minimization}
We consider a Hamiltonians $H$ as a summation of local terms, $H=\sum_{i=1}^{N} c_i L_i$, where a local Hamiltonian $L_i$ can be written as a tensor product of a few number of Pauli matrices. We denote $\bc=(c_1,c_2,...,c_N)^T$ and $\boldsymbol{L}=(L_1,L_2,...,L_N)^T$. Thus we can write $H=\bc^T\boldsymbol{L}$.

The variational quantum eigensolver works as follows. One uses an ansatz $\kets{\psi(\bt)}=U(\bt)\kets{R}$ to represent a candidate ground state. Here $\kets{R}$ usually is taken as an approximation for the ground state of $H$, e.g., Hartee-Fock state for the quantum chemistry problem.  $U(\bt)$ is an unitary operator parameterized with $\bt$. The task then is to optimize $\bt$ for some cost function. The tradition way is to minimize the energy defined as
\begin{equation}
\mathcal{E}(\bt)=\bras{\psi(\bt)}H\kets{\psi(\bt)}=\text{Tr}[\psi(\bt)H].
\end{equation}
where we have denoted $\psi(\bt)=\kets{\psi(\bt)}\bras{\psi(\bt)}$. By designing, VQE based on minimizing energy is suitable for finding ground state of a Hamiltonian.

The optimization is completed with a hybrid quantum-classical algorithm. The quantum processor prepares $\psi(\bt)$ and performs measurements to evaluate $\mathcal{E}(\bt)$, which can be reduced into $\mathcal{E}(\bt)=\bc^T\boldsymbol{\mathcal{L}}(\bt)$, where $\boldsymbol{\mathcal{L}}(\bt)=\text{Tr}(\psi(\bt)\boldsymbol{L})$.
Here, quantum average of each component of $\boldsymbol{L}$ corresponds to a joint measurement on multiple qubits.
The classical computer updates parameters $\bt$ according to received data from the quantum processor, e.g., using gradient descent or gradient-free methods.


\subsection{VQE by variance minimization}
As the name indicates, variational quantum eigensolver should aim for solving eigenstates. However, VQE, based on minimizing energy, is prone to find only an eigenstate with the lowest energy, namely the ground state. Other VQEs have been developed for solving excited states. Essentially, those VQEs are realized by minimizing energy in a subspace, which is enforced with symmetry, or by eliminating space of lower energy states with an orthogonality condition.

Can variational quantum eigensolver directly solve excited states?
An answer for this is to design cost function that can assign the same cost to all eigenstates and higher cost for other states. A natural choice is energy variance, which is zero only for eigenstates and positive for others. By minimizing energy variance, one then can find the eigenstates of a Hamiltonian. We call this as variance-VQE~(VVQE), and energy-VQE(EVQE) for VQE that minimizes the energy.

Let us formulate variance-VQE. The energy variance of Hamiltonian $H$ with a wavefunction $\psi(\bt)$ can be written as,
\begin{eqnarray}\label{eq:energy_variance}
\varDelta(\bt)&\equiv&\left<H^2\right>_{\bt}-\left<H\right>_{\bt}^2 \ge 0.
\end{eqnarray}
where $\left<\ast\right>_{\bt}=\bras{\psi(\bt)}\ast\kets{\psi(\bt)}$.
It is useful to introduce the quantum covariance matrix~\cite{bairey_19,chertkov_18,Qi2019determininglocal} defined as
\begin{eqnarray}
\mathcal{G}_{ij}(\bt)&=&\left<L_iL_j\right>_{\bt}-\left<L_i\right>_{\bt}\left<L_j\right>_{\bt}\nonumber \\
&=&\text{Tr}\left[\psi(\bt)L_iL_j\right]-\text{Tr}\left[\psi(\bt)L_i\right]\text{Tr}\left[\psi(\bt)L_j\right].
\end{eqnarray}
Then, the energy variance can be expressed as
\begin{equation}
\varDelta(\bt)=\bc^T\mathcal{G}(\bt)\bc \ge 0
\end{equation}
The nonnegative energy variance is consisted with that $\mathcal{G}(\bt)$ by definition is semi-positive matrix. Note that each element of $\mathcal{G}(\bt)$ can be obtained on a quantum computer. Then,
the energy variance can be evaluated based on the quantum covariance matrix.

Minimization energy variance is carried with gradient free or gradient descents. Gradient descent method requires to calculate  the gradient of energy variance, which is:
\begin{equation}\label{eq:gradient_variance}
\frac{\partial\varDelta(\bt)}{\partial\bt}=\bc^T\frac{\partial\mathcal{G}(\bt)}{\partial\bt}\bc.
\end{equation}
Each element $\frac{\partial\mathcal{G}_{ij}}{\partial\bt}$ can be obtained using numeral differential or the shift rule~\cite{schuld_19}. The gradient descent updates $\bt$ as
\begin{equation}\label{eq:gd}
\bt^{t}=\bt^{t-1}-\eta \frac{\partial\varDelta(\bt^{t-1})}{\partial\bt}.
\end{equation}

Since $\mathcal{G}(\bt)$ should be evaluated with $O(N^2)$ elements, calculating energy variance costs $O(N^2)$, which is larger than a cost of $O(N)$ for calculating energy.
As a result, a direct gradient descent for variance-VQE can be more resource-costing than energy-VQE, as the optimization process is expected to have multiple runs.
Nevertheless, we can use Hamiltonian sampling to reduce this cost~\cite{sweke_19}, which estimates the energy variance and its gradient by sampling only a portion of the quantum covariance matrix. In Sec.~\ref{sec:hamiltonian sampling}, we will give a stochastic gradient descent algorithm for optimizing variance-VQE by Hamiltonian sampling~\cite{sweke_19}, which can be efficient under a  minimal sampling rate and thus can reduce the overload significantly.
Besides, one may reduce the overload of evaluating the energy variance using some smarter strategies that can  simultaneously access a set of quantum covariance matrix elements~\cite{huggins2019efficient,cotler_20}.

\section{Variance-VQE for excited states}\label{sec:excited_states}
As all eigenstates always have zero variance, optimizing variance-VQE by minimizing energy variance may lead to any eigenstate. Thus, variance-VQE can be used for  obtaining excited states. In this section, we propose different approaches for solving excited states of molecule Hamiltonians with variance VQE. We first show that the cost function variance-VQE can have many global minimums of zero variance, corresponding to different
excited states. However, this approach can be inefficient if a specified eigenstate is required. To overcome this, we propose variance-VQE that solves a set of excited states jointly, using a set of orthogonal wavefunction ansatz. We further demonstrate that optimization of a linear combination of energy and variance can be more efficient than optimizing  energy or variance alone.

For demonstration, we use the hydrogen molecule $H_2$ as an example. We would consider a more complicated system, $H_4$, in Sec.~\ref{sec:hamiltonian sampling}. All Hamiltonians are calculated with \emph{OpenFermion}~\cite{OpenFermion}, and the numeral simulation for variance VQE is conducted with the HiQ simulator framework~\cite{HiQ}. Under STO-3G basis, we use four qubits to describe the Hamiltonian of $H_2$. An UCC ansatz is taken with trotter step $k=1$~(see Appendix.~\ref{append:ucc_ansatz}), and there are $9$ parameters~(6 for single excitations and 3 for double excitations). The reference state is chosen as $\kets{0011}$, or others in the subspace of two electrons.

\subsection{Single ansatz with multiple global minimums}
For the variance VQE, a criteria for being eigenstates is that energy variance $\varDelta(\bt)=0$.
If a wavefunction ansatz can represent all eigenstates, then it is expected that there are many solutions of $\bt$ that $\varDelta(\bt)=0$. In practice, as high energy eigenstates usually have higher complexity, an ansatz may only capture some eigenstates. In the optimization, one can select those minimums that are close to zero. This is indeed one advantage of variance-VQE: one can verify if an eigenstate is obtained by checking whether the variance is zero~\cite{kokail_19}.
\begin{figure}
	\includegraphics[width=0.8\columnwidth]{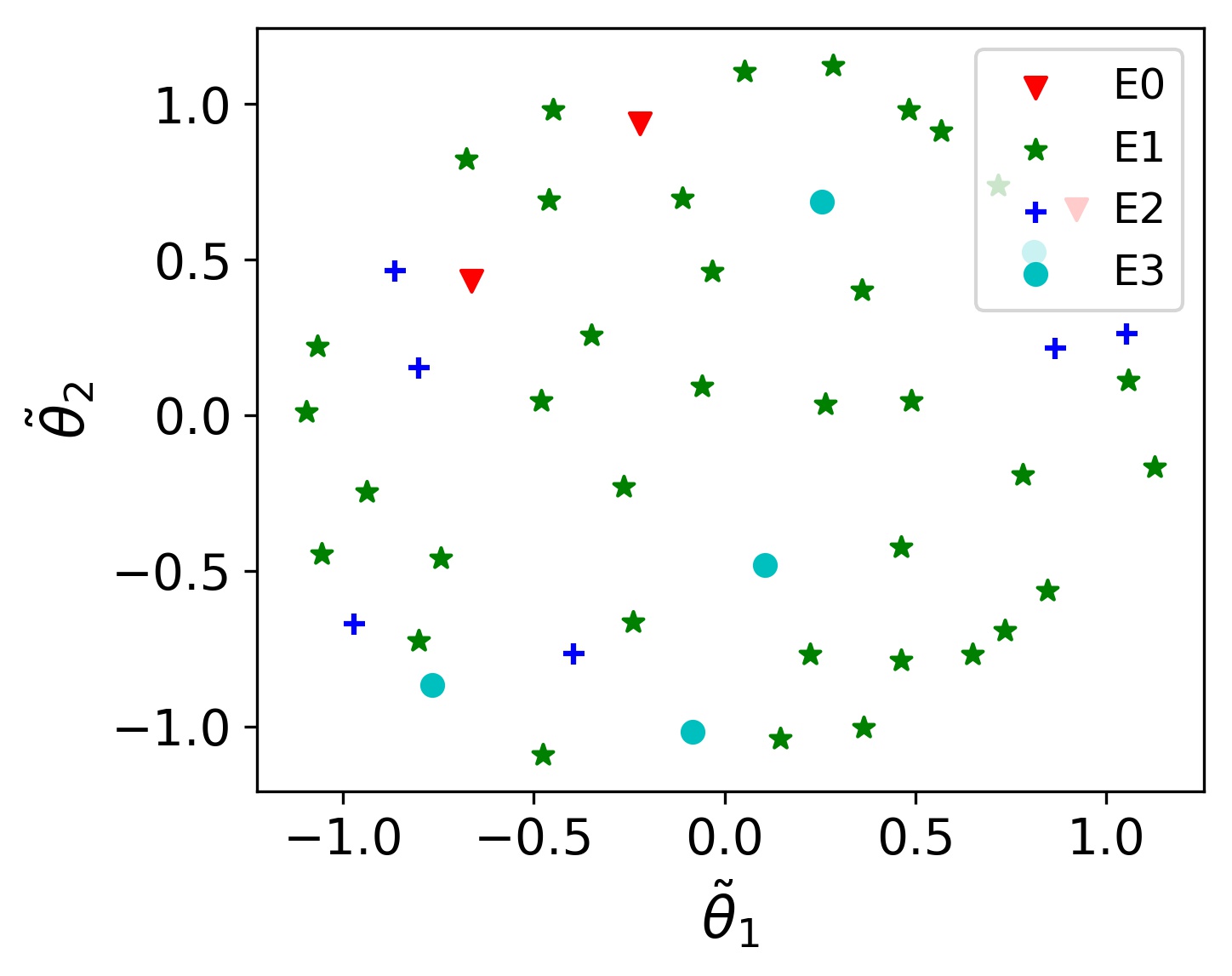}
	\caption{Distribution of global minimums in the parameter space $\bt$, where the high dimension $\bt$ is visualized on a plane, using multi-dimensional scaling method. $E0,E1,E2,E3$ correspond to energy levels in an ascending order.}
	\label{global_minimums}
\end{figure}
A direct way to calculate excited states with variance VQE is as follows. First, chose an ansatz $\kets{\psi(\bt)}$ that can express considered excited states. For quantum chemistry, we can use some modified UCC ansatz with enough single-particle and double-particle excitations~(see Appendix.~\ref{append:ucc_ansatz}). Second, minimizing $\varDelta(\bt)$ with randomly choice of initial parameters $\bt_0\in [0,2\pi]^{K}$. Third, select minimums of $\varDelta(\bt)$ that are close to zero. Lastly, with optimized parameters, we can calculate their corresponding energies.

As for demonstration, we calculate spectrum of $H_2$. With uniformly sampled initial $\bt$, we optimize the energy variance with default optimizer in \emph{Scipy}. All obtained minimums are almost zero~(less than $10^{-8}$), and their corresponding energies are eigen-energies successfully. However, we find the chance to be a given eigenstate varies, e.g., solutions to ground states are far less than excited states. To illustrate this, we visualize the distribution of solutions in the parameter space, as seen in Fig.~\ref{global_minimums}, by projecting the 9-dimensional parameters onto a plane, using a multidimensional scaling method that can preserve well information of distance. Note we use a distance metric $d(\bt,\bt')=\sum_{i}\cos(\theta_i-\theta'_i)$ as $\theta$ is an angle that $\theta_1=\theta_2~\text{mod}~2\pi$. This indicates that the ground state is relatively hard to find in this method. It is also observed that multiple points of parameters may correspond to the same excited state. This can be understand as there exists many $U(\bt)$ that can transform $\kets{R}$ to the same target state. The above conclusion is the same for other reference states, e.g., $\kets{0101}$ and $\kets{0110}$.

\subsection{A set of orthogonal ansatzs}
While variance-VQE can be optimized to obtain all eigenstates with a single ansatz with different optimized parameters, it can be hard to get the ground state as it requires choosing a proper initial parameter. Here we develop another method, using a set of orthogonal ansatz wavefunctions.

The orthogonality of ansatzs for different eigenstates can be easily enforced by using $\kets{\psi_n(\bt)}=U(\bt)\kets{R_n}$, where $\{\kets{R_n}\}$ are orthogonal to each other. Note that all ansatz uses the same $U(\bt)$ and the orthogonality of $\{\kets{\psi_n(\bt)}\}$ only holds under the same $\bt$. To optimize $\bt$, we use a cost function that is equal-weighting summation of all energy variances, which can be written as,
\begin{equation}\label{eq:cost_all_zero_variance}
\mathcal{C}_\text{var}(\bt)=\sum_{n=0}^{k-1}w_n\varDelta_n(\bt),
\end{equation}
where  $\varDelta_n(\bt)$ is the energy variance of $H$ for the state $\kets{\psi_n(\bt)}$, and $w_n=1/k$. The formula is similar to that of subspace-search VQE, which requires a specified weighting $\{w_n\}$ that enforces an ordering of eigenstates.  We note that Eq.~\eqref{eq:cost_all_zero_variance} has been used in Ref.~\cite{pollmann_16} to calculate all eigenstates of quantum many-body system with unitary matrix product ansatz.
\begin{figure}[H]
	\includegraphics[width=1\columnwidth]{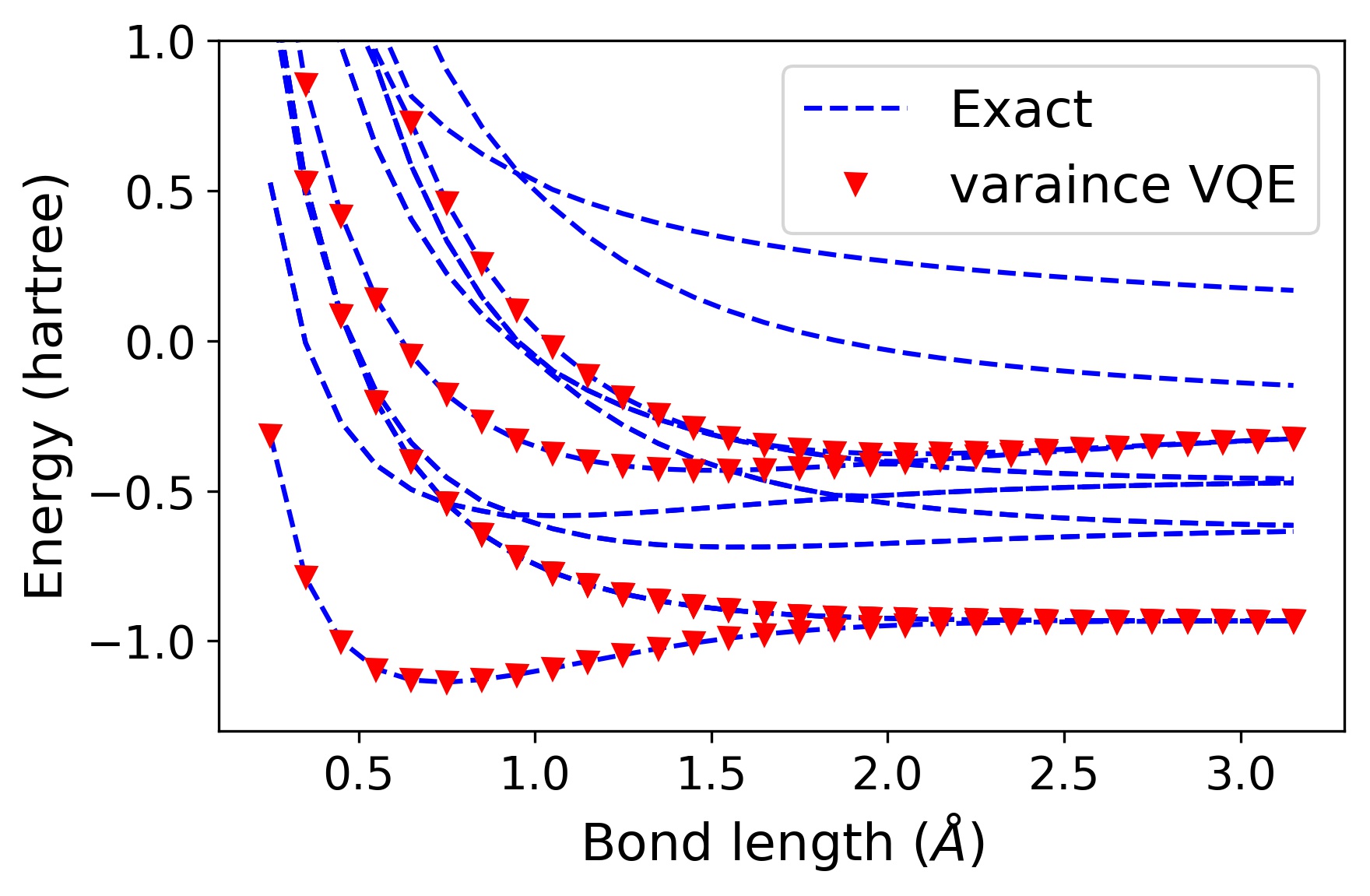}
	\caption{Spectrum of $H_2$ with bond length. Blue dash lines are results from exact diagonalization in the whole Hilbert space. Red markers are obtained from variance-VQE in a subspace of two electrons that $H_2$ is electronic neutral.}
	\label{fig:spectrum}
\end{figure}

\begin{figure}[H]
	\includegraphics[width=1\columnwidth]{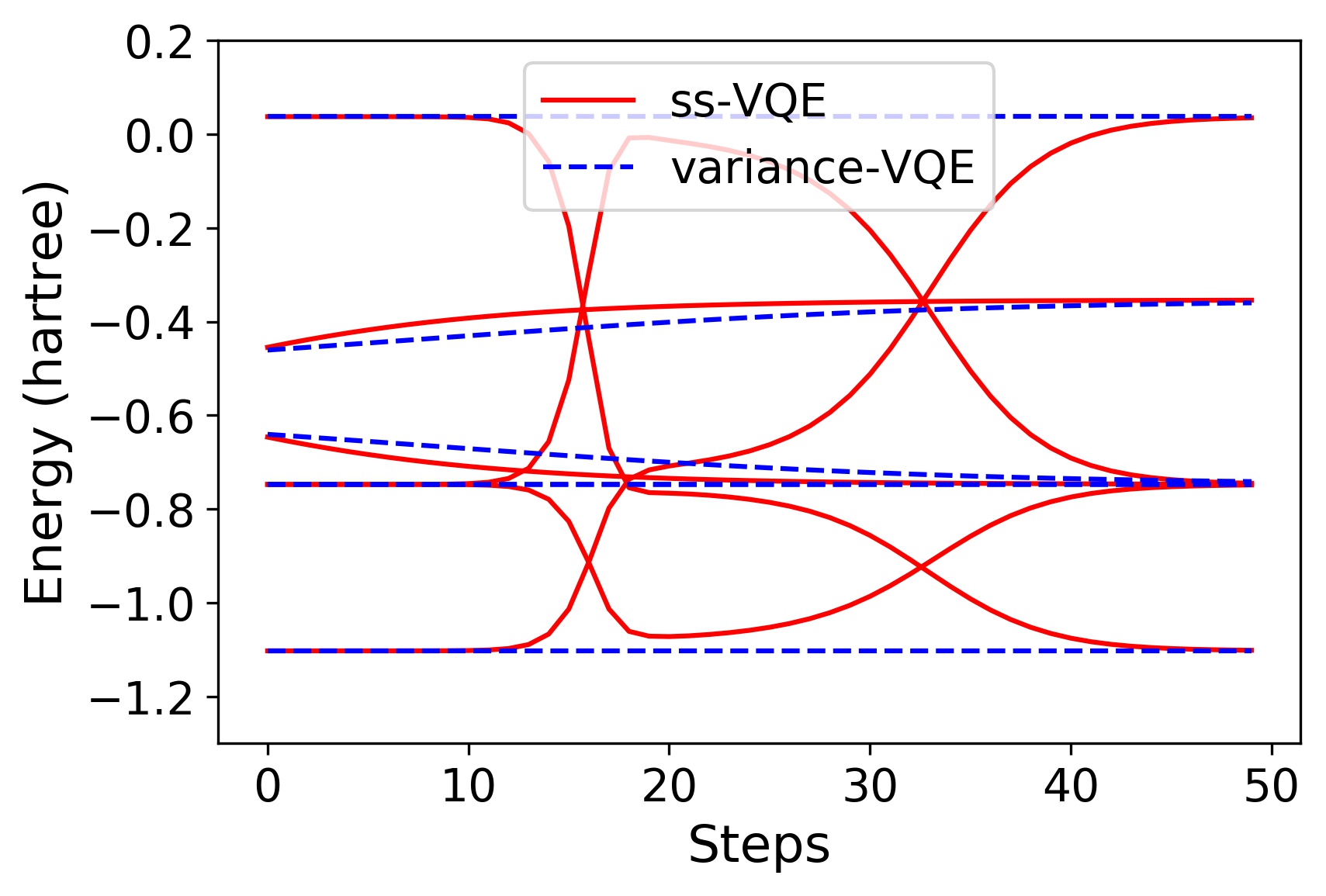}
	\caption{Evolution of energy levels in the process of optimization.
		Two methods are compared: the subspace-search VQE and the variance-VQE.The bond length for $H_2$ is $\lambda=1$. }
	\label{fig:Evolution_spectrum}
\end{figure}
While other methods usually calculate excited states one by one, this method incorporates all into a single cost function to optimize. In spirit, it can be viewed as a multi-task learning~\cite{multi_task_learning}. As those tasks are closely related, solving them in a package may benefit each other.

Solving excited states then can be done by optimizing $\mathcal{C}_\text{var}(\bt)$ with gradient descent, \begin{equation}\label{eq:gd}
\bt^{t}=\bt^{t-1}-\eta \frac{\partial\mathcal{C}_{\text{var}}(\bt^{t-1})}{\partial\bt}.
\end{equation}
Other gradient-free methods are also possible. We present the numeral results for $H_2$. Reference states are computational basis on a subspace with 2 particles, which has $6$ states. As the UCC ansatz is particle-conserving, the wavefunction ansatz is in the subspace of 2 particles. The energy potential curve fits perfectly to that by exact diagonalization, as seen in Fig.~\ref{fig:spectrum}.

It is noted that evaluation of the cost function Eq.~\eqref{eq:cost_all_zero_variance} can be more efficient with ancillary qubits, which is given in the Appendix~\ref{appendix:eval_cost_function}. This is useful when lots of eigenstates are needed,  where a direct summation can be impractical. For instance, there is an exponentially large number of terms when all eigenstates are needed.

\subsubsection*{Comparison with subspace-search VQE}
It is interesting to compare the result of variance-VQE with orthogonal ansatz with that of subspace-search VQE~\cite{nakanishi_19}.
The subspace-search VQE focuses on a subspace of eigenstates function and optimizes a number of $k$ low-energy eigenstates simultaneously. The cost function is a weighted combination of all energies,
\begin{equation}\label{eq:cost_all}
\mathcal{C}(\bt)=\sum_{n=0}^{k-1} w_n\mathcal{E}_n(\bt)=\sum_{n=0}^{k-1} w_n \text{Tr}\left[\psi_n(\bt)H\right],
\end{equation}
where $\kets{\psi_n(\bt)}=U(\bt)\kets{n}$. The weightings satisfy $w_0>w_1>w_2>...>w_{k-1}$, such that energies at optimized $\bt^*$ will satisfy $\mathcal{E}_0(\bt^*)\le\mathcal{E}_1(\bt^*)\le...\le\mathcal{E}_{k-1}(\bt^*)$~(see a proof in Ref.~\cite{nakanishi_19}). It is noted that the weightings can be learned, as in the Ref.~\cite{liu_19_gibbs}, which self-adjusts the weightings when variational preparing Gibbs state for a quantum system.

A problem for the subspace-search VQE is that the order of eigenstates needs to be given. It is possible that the optimization may go a long way in order to make all eigenstates rest in a given order, as revealed in Fig.~\ref{fig:Evolution_spectrum}. This can increase the complexity of the optimization. To illuminate this, we compare the evolution of energies in the optimization process for both subspace-search VQE and variance-VQE. The ordering of eigenstates in subspace-search VQE is fixed by reference states $\kets{0011},\kets{0101},\kets{1001},\kets{0110},\kets{1010},\kets{1100}$,which give to ascending order of energies after optimization.  As seen in Fig.~\ref{fig:Evolution_spectrum}, initialized with the same parameters and thus the same energies, evolutions of energies for subspace-VQE will have complicated trajectories toward  convergence. In contrast, all orthogonal ansatz wavefunctions flow directly to nearest eigenstates in the variance-VQE. We also check other ordering of eigenstates for subspace-search VQE, and find the same result. In this aspect, variance-VQE can be easier to optimize.


\subsection{Minimizing a combination of energy and variance}

For the orthogonal ansatz, if we only chose a subset of the Hilbert space or subspace with a fixed number, then variance-VQE can not promise that the solutions are lowest energy states. Just as the previous approach using one ansatz, the final result depends on the initial parameter and the optimization process. On the other hand, for a cost function with an equal-weighting of energies for orthogonal ansatz, it can get the lowest total energy, but can not promise each optimized state is an eigenstate. While subspace-search VQE can solve this problem, the optimization can be not efficient by assigning an order for ansatzs as eigenstates. Here, we demonstrate that a simple combination of energies and energy variances, can make the best of both  to find low energy excited states.
\begin{figure}
	\includegraphics[width=1\columnwidth]{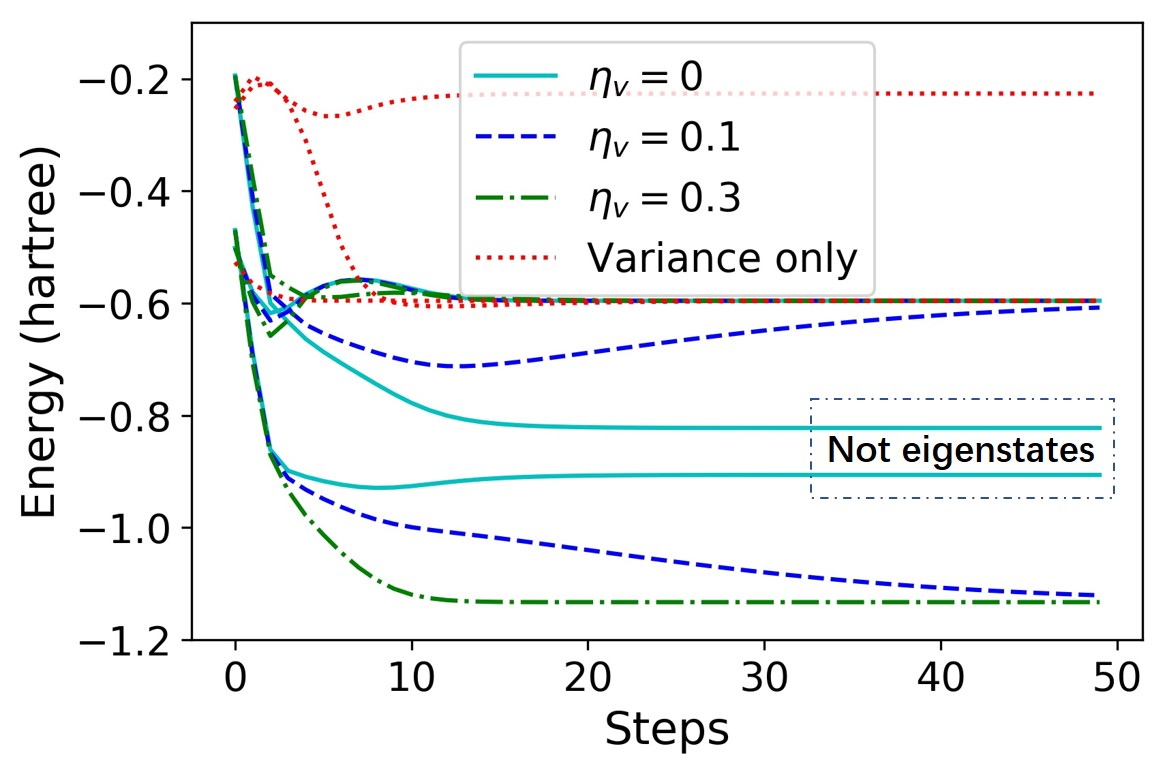}
	\caption{(Color online) Optimization processes for mixed cost function with different mixing factor $\eta_v$ of energy variances. Three orthogonal ansatz are used. The bond length for $H_2$ is $\lambda=0.8$. }
	\label{fig:mixed_VQE}
\end{figure}
The cost function as a combination of energies and energy variances can be written as,
\begin{equation}\label{eq:mixed_cost}
\mathcal{C}_\text{mix}(\bt)=\sum_{n=0}^{k-1}\mathcal{E}_n(\bt)+\eta_v\sum_{n=0}^{k-1}\varDelta_n(\bt).
\end{equation}
Here, all weights are the same, and $\eta_v$ a mixer factor. By minimizing $\mathcal{C}_\text{mix}(\bt)$, the first term of Eq.~\eqref{eq:mixed_cost} flows to low energy subspace while the second term makes sure that optimized states are eigenstates.  A gradient descent process for optimizing  $\mathcal{C}_\text{mix}(\bt)$ is,
\begin{equation}\label{eq:gd_hybrid}
\boldsymbol{\theta}^{t}=\boldsymbol{\theta}^{t-1}-\eta_A\sum_n \frac{\partial\mathcal{E}_n(\boldsymbol{\theta}^{t-1})}{\partial{\boldsymbol{\theta}}}-\eta_A\eta_v\sum_n \frac{\partial\varDelta_n(\boldsymbol{\theta}^{t-1})}{\partial\boldsymbol{\theta}}.
\end{equation}
In this regard, the optimization includes gradient information of both energy and energy variance, and $\eta_v$ plays a role  in adjusting the step size between two.
In Fig.~\ref{fig:mixed_VQE}, we show that the result of mixed VQE with different $\eta_v$, and it can be seen that nonzero $\eta_v$ is critical to get the right eigenstates, and increasing $\eta_v$ can raise the efficiency for optimization. Moreover, if only energy variance is used in the cost function, the solution may not be the lowest energy eigenstates. Thus, the result demonstrates that a combination of energy and energy variance can be useful and efficient for VQE to solve low-energy excited states.

\section{Hamiltonian sampling}\label{sec:hamiltonian sampling}
By now, we have demonstrated that variance-VQE can efficiently solve excited states for a Hamiltonian. However, the overload of calculating variance and its gradient descent is massive since it scales with the number of Hamiltonian terms as $O(N^2)$. In this section, we propose stochastic gradient descent for variance-VQE by Hamiltonian sampling, which can significantly reduce the overload.

The Hamiltonian sampling randomly chose some components of $\bc$ (set $c_i=0$ if $c_i$ if not chosen), which is denoted as $\tilde{\bc}$. Then, the estimated energy variance is
\begin{eqnarray}
\tilde{\varDelta}(\bt)=\frac{|\bc|^2}{|\tilde{\bc}|^2}\tilde{\bc}^T\mathcal{G}(\bt)\tilde{\bc}.
\end{eqnarray}
A prefactor $\frac{|\bc|^2}{\tilde{\bc}^2}$ is added to account for the fact that coefficients for different terms of a Hamiltonian varies largely. Whenever $\tilde{c}_i=0$, elements $\mathcal{G}(\bt)_{ij}$ do not need to be evaluated. For a sampling rate $s$~(defined as a ratio that $c_i$ is sampled), the number of elements in $\mathcal{G}(\bt)_{ij}$ required to be calculated can be reduced by a factor $s^2$.

With a Hamiltonian sampling, the gradient can be estimated as $\frac{\partial\tilde{\varDelta}(\bt)}{\partial\bt}$, which shall have a distribution due to sampling. With this gradient, a stochastic gradient descent algorithm can be applied for optimizing variance-VQE, which updates parameters as
\begin{equation}\label{eq:gd}
\bt^{t}=\bt^{t-1}-\eta \frac{\partial\tilde{\varDelta}(\bt^{t-1})}{\partial\bt}.
\end{equation}

We apply the stochastic gradient descent for solving eigenstates of $H_2$ and $H_4$ with variance-VQE. The molecule $H_4$ is investigated in a trapezoidal structure. Under sto-6g basis, we use a Hamiltonian of $6$ qubits to describe $H_4$. The UCC ansatz is chosen as $k=1$ with $21$ parameters, including $15$ for single-particle excitations and $6$ for double-particle excitations~(we only consider double-particle excitations from occupied orbitals to empty orbitals). Note that there are more than one hundred terms for the Hamiltonian, which means that Hamiltonian sampling is badly demanded to evaluate the variance.

\begin{figure*}\centering
	\includegraphics[width=15cm]{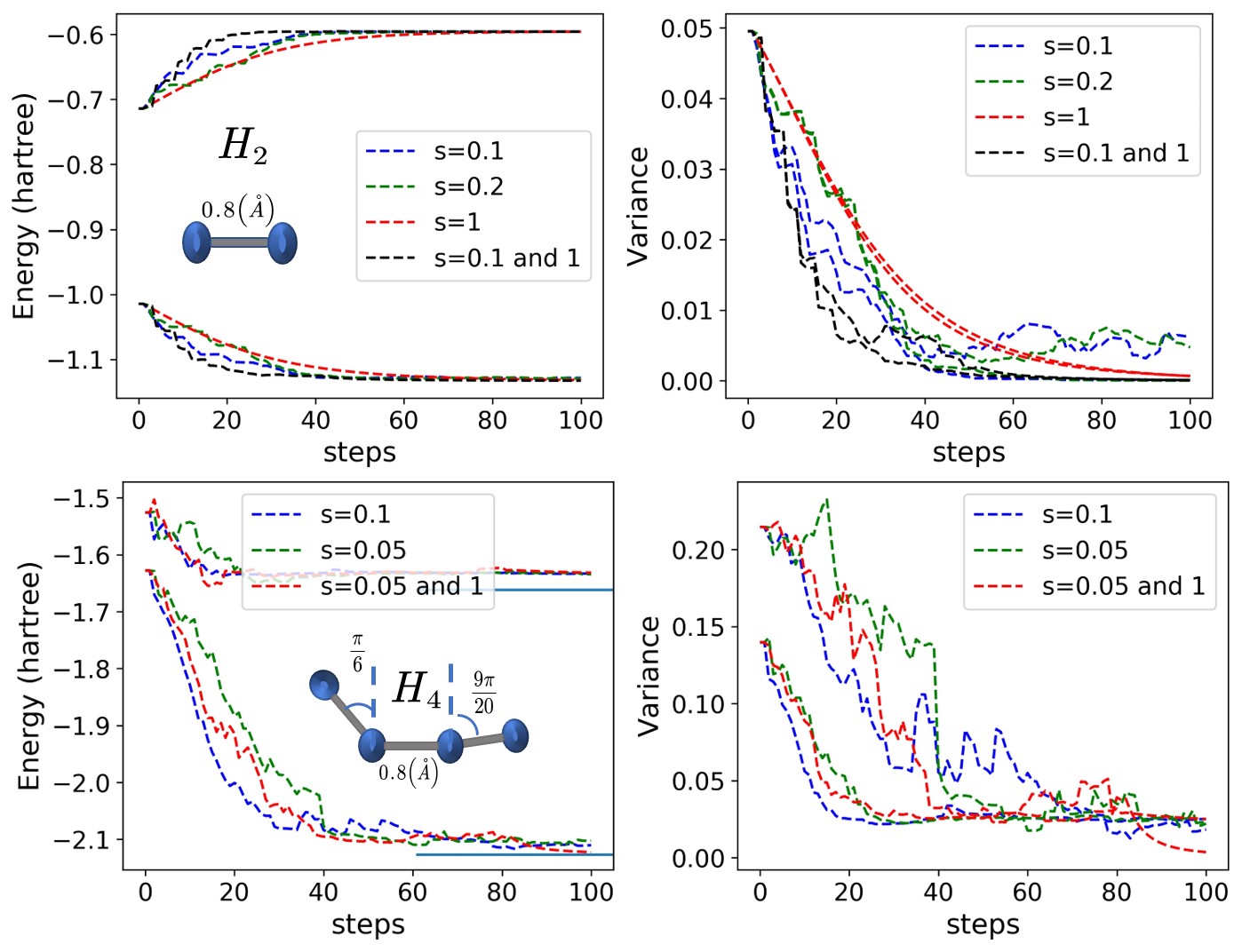}
	\caption{(Color online) Optimization of variance-VQE for $H_2$ (upper row) and $H_4$~(down row) with different Hamiltonian sampling rates $s$. Sampling rates are adjustable, for instance, $s=0.1~\text{and}~1$ means $s$ changes from $0.1$ to $1$ at the later stage of optimization. The learning rate is $0.05$ for all cases.}
	\label{fig:ham_sampling}
\end{figure*}

Numeral simulation results are displayed in Fig.~\ref{fig:ham_sampling} for solving two eigenstates with orthogonal ansatz. For $H_2$, it shows that Hamiltonian sampling with a small sampling rate $s=0.1,0.2$ can have comparable convergent behavior with the case without sampling. However, the variance ceases to converge at small values, due to the fluctuation of energy for a subsystem even when the whole system has zero energy variance. To solve this issue, we can turn to no sampling (sampling rate $s=1$) at the late stage of optimization. For $H_4$, similar phenomena can be observed. It is noted that only one ansatz reaches the true eigenstate, while the other fail~(a local minimum or the ansatz lacks the capacity to express this eigenstate), which can be seen from the nonzero energy variance. This again verifies that variance-VQE can self-verify whether an eigenstate is obtained.

We can explain why Hamiltonian sampling fails to work around the optimized point of parameters, where gradients by Hamiltonian sampling can give meaningless direction for optimization. This is because energy variance is not additive: if $H=H_A+H_B$ and the wavefunction is $\kets{\psi}$, then the variance of $H$ is not a summation of variance of $H_A$ and variance of $H_B$. For example, if $\kets{\psi}$ is an eigenstate of $H$, $\kets{\psi}$ will not be an eigenstate of $H_A$ or $H_B$, if $A$ and $B$ are entangled. As a result, even at the optimized parameter of zero variance, gradients by Hamiltonian sampling can be finite.
Such a property is in sharp contrast to the case of machine learning. Machine learning usually uses a set of independent and identically distributed samples, and the cost function is a summation of independent contributions from all samples. Thus, gradients can be estimated with a batch of samples, giving stochastic gradient descents with well convergence behavior~\cite{Goodfellow-et-al-2016}.

From the above discussion, a practical strategy is to use Hamiltonian sampling with a small sampling rate to optimize the variance at a small value, then turn to a large sampling rate or no sampling. Also, one may then use energy-VQE.  We note that stochastic gradient descent with few shots of measurements can further reduce the overload~\cite{sweke_19,Kubler2020adaptiveoptimizer}, which leaves for further investigation. 


\section{Summary}\label{sec:summary}
In summary, we have proposed the variance-VQE that optimizes the energy variance to obtain eigenstates for a Hamiltonian. Compared with VQE that minimizes the energy which is prone to find ground state, the variance-VQE naturally is suitable for solving  and self-verifying arbitrary eigenstates. We have adopted different strategies to solve excited states with the variance-VQE, using molecules $H_2$ and $H_4$ as examples. Remarkably, it has been shown that optimizing a set of orthogonal ansatzs with the same parameterized circuit can be very efficient for calculating a set of eigenstates. Also, we have demonstrated that minimizing a linear combination of energy and variance can outperform the case of optimizing either energy or variance alone. Moreover, we have proposed stochastic gradient descent for minimizing the variance by Hamiltonian sampling, using only a few terms of the Hamiltonian to estimate the variance and its gradient. It has been shown that the variance can decrease quickly under a small sampling rate. While numeral simulations suggested that the variance fails to converge, Hamiltonian sampling can be useful at the early stage to locate the zone of optimized parameters. Our work has demonstrated that variance-VQE can be useful and practical for variational solving excited states for a quantum system on a quantum computer.

\begin{acknowledgments}
	This work was supported by the Key-Area Research and Development Program
	of GuangDong Province (Grant No. 2019B030330001), the National Key Research
	and Development Program of China (Grant No. 2016YFA0301800), the National Natural Science Foundation of China (Grants No. 91636218 and
	No. U1801661), the Key Project of Science and Technology of Guangzhou (Grant No. 201804020055).
\end{acknowledgments}

\appendix{
\section{UCC ansatz}\label{append:ucc_ansatz}
The UCC ansatz is widely used in the field of quantum chemistry, as it can represent a parametrized wavefunction of electronic structure of a molecule efficiently. The parametrized wavefunction can be written as,
\begin{equation} \label{eq:ucc}
\kets{\psi(\boldsymbol{\theta})}=e^{T-T^\dagger}\kets{R}.
\end{equation}
Here, $T=T_1+T_2$ usually includes single-particle excitations $T_1$ and double-particle excitations $T_2$.
We adopt a notation that does not distinguish occupied and empty states in the reference state $\kets{R}$ (which typically is Hartree-Fock state). Then $T_1$ and $T_2$ can be expressed as,
\begin{eqnarray}\label{eq:uccsd}
&&T_1=\sum_{pq} \theta^p_q c_p^\dagger c_q \nonumber \\
&&T_2=\sum_{pqrs}\theta^{pq}_{rs}c_p^\dagger c_q^\dagger c_r c_s.
\end{eqnarray}

To implement Eq.~\eqref{eq:ucc}, a trotterization of $e^{T-T^\dagger}$ is required. However, $\{t^p_q\}$
and $\{t^{pq}_{rs}\}$ are parameters that should be optimized at first. This makes a formula as Eq.~\eqref{eq:ucc} can not be directly implemented on a quantum computer. Due to the variational flexibility, one can use a few Trotter step (even a single Trotter step may work good). The case of a single Trotter step can be written as~\cite{grimsley_19},
\begin{equation}
\kets{\psi(\boldsymbol{\theta})}=\prod_{pq}e^{\hat{t}_{pq}}\prod_{pqrs}e^{\hat{t}_{pqrs}}\kets{R},
\end{equation}
where $\hat{t}_{pq}=\theta^p_q(c_p^\dagger c_q-c_q^\dagger c_p)$ and
$\hat{t}_{pqrs}=\theta^{pq}_{rs}(c_p^\dagger c_q^\dagger c_r c_s-c_s^\dagger c_r^\dagger c_p c_q)$.
For more Trotter steps, one can use~\cite{grimsley_19},
\begin{equation}
\kets{\psi(\boldsymbol{\theta})}=\prod_{k=1}^{n}\prod_{pq}e^{\hat{t}^k_{pq}}\prod_{pqrs}e^{\hat{t}^k_{pqrs}}\kets{R}.
\end{equation}	
}
\section{Efficient evaluation of cost function}\label{appendix:eval_cost_function}
While the goal is to solve a number of $k$ excited states, evaluation of the cost function (the total energy variance)
is time consuming for large $k$, since it should calculate energy variance one by one. For instance, for a system of N qubits, the total number of eigenstates is $2^N$. Here, we discuss how to efficient evaluate $\mathcal{C}_\text{var}(\bt)$, similar to Ref.~\cite{pollmann_16}. Let us first consider $k=2^N$. Using
\begin{equation}
I = \sum_{n=0}^{2^N-1}\kets{\psi_n(\bt)}\bras{\psi_n(\bt)},
\end{equation} Eq.~\eqref{eq:cost_all_zero_variance} can be rewritten as
\begin{eqnarray}
&&\mathcal{C}_\text{var}(\bt) \nonumber \\
&&=Tr[H^2\sum_{n=0}^{2^N-1}\kets{\psi_n(\bt)}\bras{\psi_n(\bt)}] - \sum_{n=0}^{2^N-1}\bras{\psi_n(\bt)}H\kets{\psi_n(\bt)}^2 \nonumber \\
&&=Tr[H^2]-\sum_{n=0}^{2^N-1}\bras{\psi_n(\bt)}H\kets{\psi_n(\bt)}^2.
\end{eqnarray}
The first term equals to $|\boldsymbol{c}|^2$. The second still involves a summation of exponential number terms. However, we can sum up them in a quantum parallel way.
By preparing a state,
\begin{eqnarray} \label{eq:large_state}
\kets{\phi(\bt)}&=&U(\bt) \otimes U(\bt) \otimes I \frac{1}{\sqrt{2^N}}\sum_{n=0}^{2^N-1}\kets{n}\kets{n}\kets{n} \nonumber \\
&=&\sum_{n=0}^{2^N-1} \kets{\psi_n(\bt)}\kets{\psi_n(\bt)}\kets{n},
\end{eqnarray}
then it can be verified that
\begin{equation}\label{eq:large_state}
\bras{\phi(\bt)} H\otimes H\otimes I \kets{\phi(\bt)}
=\frac{1}{2^N}\sum_{n=0}^{2^N-1}\bras{\psi_n(\bt)}H\kets{\psi_n(\bt)}^2. \nonumber
\end{equation}
This indicates that the cost function can be efficiently calculated, at the price of increasing the number of qubits form $N$ to $3N$, which makes a good deal since it can sum up an number of $2^N$ terms.

For general $k<2^N$, we can adopt similar ideas.
We introduce $M$~(requiring $2^M\geq k>2^{M+1}$) ancilla qubits to prepare
\begin{eqnarray}
\kets{\varphi(\bt)}=U(\bt) \otimes I \frac{1}{\sqrt{k}}\sum_{n=0}^{k-1}\kets{n}\kets{n}
= \sum_{n=0}^{k-1} \kets{\psi_n(\bt)}\kets{n}, \nonumber
\end{eqnarray}
then the first part of Eq.~\eqref{eq:cost_all_zero_variance} is evaluated as
\begin{equation}
k\bras{\varphi(\bt)}H^2\kets{\varphi(\bt)}.
\end{equation} To evaluate the second part of Eq.~\eqref{eq:cost_all_zero_variance}, the summation in Eq.~\eqref{eq:large_state} should be replace by $k-1$. The total number of qubits is $N+N+M=2N+M$.
%


\begin{thebibliography}{51}%
	\makeatletter
	\providecommand \@ifxundefined [1]{%
		\@ifx{#1\undefined}
	}%
	\providecommand \@ifnum [1]{%
		\ifnum #1\expandafter \@firstoftwo
		\else \expandafter \@secondoftwo
		\fi
	}%
	\providecommand \@ifx [1]{%
		\ifx #1\expandafter \@firstoftwo
		\else \expandafter \@secondoftwo
		\fi
	}%
	\providecommand \natexlab [1]{#1}%
	\providecommand \enquote  [1]{``#1''}%
	\providecommand \bibnamefont  [1]{#1}%
	\providecommand \bibfnamefont [1]{#1}%
	\providecommand \citenamefont [1]{#1}%
	\providecommand \href@noop [0]{\@secondoftwo}%
	\providecommand \href [0]{\begingroup \@sanitize@url \@href}%
	\providecommand \@href[1]{\@@startlink{#1}\@@href}%
	\providecommand \@@href[1]{\endgroup#1\@@endlink}%
	\providecommand \@sanitize@url [0]{\catcode `\\12\catcode `\$12\catcode
		`\&12\catcode `\#12\catcode `\^12\catcode `\_12\catcode `\%12\relax}%
	\providecommand \@@startlink[1]{}%
	\providecommand \@@endlink[0]{}%
	\providecommand \url  [0]{\begingroup\@sanitize@url \@url }%
	\providecommand \@url [1]{\endgroup\@href {#1}{\urlprefix }}%
	\providecommand \urlprefix  [0]{URL }%
	\providecommand \Eprint [0]{\href }%
	\providecommand \doibase [0]{http://dx.doi.org/}%
	\providecommand \selectlanguage [0]{\@gobble}%
	\providecommand \bibinfo  [0]{\@secondoftwo}%
	\providecommand \bibfield  [0]{\@secondoftwo}%
	\providecommand \translation [1]{[#1]}%
	\providecommand \BibitemOpen [0]{}%
	\providecommand \bibitemStop [0]{}%
	\providecommand \bibitemNoStop [0]{.\EOS\space}%
	\providecommand \EOS [0]{\spacefactor3000\relax}%
	\providecommand \BibitemShut  [1]{\csname bibitem#1\endcsname}%
	\let\auto@bib@innerbib\@empty
	\bibitem [{\citenamefont {Yung}\ \emph {et~al.}(2014)\citenamefont {Yung},
		\citenamefont {Casanova}, \citenamefont {Mezzacapo}, \citenamefont {McClean},
		\citenamefont {Lamata}, \citenamefont {Aspuru-Guzik},\ and\ \citenamefont
		{Solano}}]{yung_14}%
	\BibitemOpen
	\bibfield  {author} {\bibinfo {author} {\bibfnamefont {M.~H.}\ \bibnamefont
			{Yung}}, \bibinfo {author} {\bibfnamefont {J.}~\bibnamefont {Casanova}},
		\bibinfo {author} {\bibfnamefont {A.}~\bibnamefont {Mezzacapo}}, \bibinfo
		{author} {\bibfnamefont {J.}~\bibnamefont {McClean}}, \bibinfo {author}
		{\bibfnamefont {L.}~\bibnamefont {Lamata}}, \bibinfo {author} {\bibfnamefont
			{A.}~\bibnamefont {Aspuru-Guzik}}, \ and\ \bibinfo {author} {\bibfnamefont
			{E.}~\bibnamefont {Solano}},\ }\bibfield  {title} {\enquote {\bibinfo {title}
			{From transistor to trapped-ion computers for quantum chemistry},}\ }\href
	{\doibase 10.1038/srep03589
		https://www.nature.com/articles/srep03589#supplementary-information}
	{\bibfield  {journal} {\bibinfo  {journal} {Scientific Reports}\ }\textbf
		{\bibinfo {volume} {4}},\ \bibinfo {pages} {3589} (\bibinfo {year}
		{2014})}\BibitemShut {NoStop}%
	\bibitem [{\citenamefont {Farhi}\ \emph {et~al.}(2014)\citenamefont {Farhi},
		\citenamefont {Goldstone},\ and\ \citenamefont {Gutmann}}]{farhi_14}%
	\BibitemOpen
	\bibfield  {author} {\bibinfo {author} {\bibfnamefont {Edward}\ \bibnamefont
			{Farhi}}, \bibinfo {author} {\bibfnamefont {Jeffrey}\ \bibnamefont
			{Goldstone}}, \ and\ \bibinfo {author} {\bibfnamefont {Sam}\ \bibnamefont
			{Gutmann}},\ }\bibfield  {title} {\enquote {\bibinfo {title} {A quantum
				approximate optimization algorithm},}\ }\href
	{http://arXiv.org/abs/1411.4028} {\bibfield  {journal} {\bibinfo  {journal}
			{arXiv:1411.4028}\ } (\bibinfo {year} {2014})}\BibitemShut {NoStop}%
	\bibitem [{\citenamefont {McClean}\ \emph {et~al.}(2016)\citenamefont
		{McClean}, \citenamefont {Romero}, \citenamefont {Babbush},\ and\
		\citenamefont {Aspuru-Guzik}}]{mcclean_16}%
	\BibitemOpen
	\bibfield  {author} {\bibinfo {author} {\bibfnamefont {Jarrod~R.}\
			\bibnamefont {McClean}}, \bibinfo {author} {\bibfnamefont {Jonathan}\
			\bibnamefont {Romero}}, \bibinfo {author} {\bibfnamefont {Ryan}\ \bibnamefont
			{Babbush}}, \ and\ \bibinfo {author} {\bibfnamefont {Alán}\ \bibnamefont
			{Aspuru-Guzik}},\ }\bibfield  {title} {\enquote {\bibinfo {title} {The theory
				of variational hybrid quantum-classical algorithms},}\ }\href {\doibase
		10.1088/1367-2630/18/2/023023} {\bibfield  {journal} {\bibinfo  {journal}
			{New Journal of Physics}\ }\textbf {\bibinfo {volume} {18}},\ \bibinfo
		{pages} {023023} (\bibinfo {year} {2016})}\BibitemShut {NoStop}%
	\bibitem [{\citenamefont {O’Malley}\ \emph {et~al.}(2016)\citenamefont
		{O’Malley}, \citenamefont {Babbush}, \citenamefont {Kivlichan},
		\citenamefont {Romero}, \citenamefont {McClean}, \citenamefont {Barends},
		\citenamefont {Kelly}, \citenamefont {Roushan}, \citenamefont {Tranter},
		\citenamefont {Ding}, \citenamefont {Campbell}, \citenamefont {Chen},
		\citenamefont {Chen}, \citenamefont {Chiaro}, \citenamefont {Dunsworth},
		\citenamefont {Fowler}, \citenamefont {Jeffrey}, \citenamefont {Lucero},
		\citenamefont {Megrant}, \citenamefont {Mutus}, \citenamefont {Neeley},
		\citenamefont {Neill}, \citenamefont {Quintana}, \citenamefont {Sank},
		\citenamefont {Vainsencher}, \citenamefont {Wenner}, \citenamefont {White},
		\citenamefont {Coveney}, \citenamefont {Love}, \citenamefont {Neven},
		\citenamefont {Aspuru-Guzik},\ and\ \citenamefont
		{Martinis}}]{o’malley_16}%
	\BibitemOpen
	\bibfield  {author} {\bibinfo {author} {\bibfnamefont {P.~ J~ J}\
			\bibnamefont {O’Malley}}, \bibinfo {author} {\bibfnamefont
			{R.}~\bibnamefont {Babbush}}, \bibinfo {author} {\bibfnamefont {I.~ D}\
			\bibnamefont {Kivlichan}}, \bibinfo {author} {\bibfnamefont {J.}~\bibnamefont
			{Romero}}, \bibinfo {author} {\bibfnamefont {J.~ R}\ \bibnamefont
			{McClean}}, \bibinfo {author} {\bibfnamefont {R.}~\bibnamefont {Barends}},
		\bibinfo {author} {\bibfnamefont {J.}~\bibnamefont {Kelly}}, \bibinfo
		{author} {\bibfnamefont {P.}~\bibnamefont {Roushan}}, \bibinfo {author}
		{\bibfnamefont {A.}~\bibnamefont {Tranter}}, \bibinfo {author} {\bibfnamefont
			{N.}~\bibnamefont {Ding}}, \bibinfo {author} {\bibfnamefont {B.}~\bibnamefont
			{Campbell}}, \bibinfo {author} {\bibfnamefont {Y.}~\bibnamefont {Chen}},
		\bibinfo {author} {\bibfnamefont {Z.}~\bibnamefont {Chen}}, \bibinfo {author}
		{\bibfnamefont {B.}~\bibnamefont {Chiaro}}, \bibinfo {author} {\bibfnamefont
			{A.}~\bibnamefont {Dunsworth}}, \bibinfo {author} {\bibfnamefont {A.~ G}\
			\bibnamefont {Fowler}}, \bibinfo {author} {\bibfnamefont {E.}~\bibnamefont
			{Jeffrey}}, \bibinfo {author} {\bibfnamefont {E.}~\bibnamefont {Lucero}},
		\bibinfo {author} {\bibfnamefont {A.}~\bibnamefont {Megrant}}, \bibinfo
		{author} {\bibfnamefont {J.~ Y}\ \bibnamefont {Mutus}}, \bibinfo {author}
		{\bibfnamefont {M.}~\bibnamefont {Neeley}}, \bibinfo {author} {\bibfnamefont
			{C.}~\bibnamefont {Neill}}, \bibinfo {author} {\bibfnamefont
			{C.}~\bibnamefont {Quintana}}, \bibinfo {author} {\bibfnamefont
			{D.}~\bibnamefont {Sank}}, \bibinfo {author} {\bibfnamefont {A.}~\bibnamefont
			{Vainsencher}}, \bibinfo {author} {\bibfnamefont {J.}~\bibnamefont {Wenner}},
		\bibinfo {author} {\bibfnamefont {T.~ C}\ \bibnamefont {White}}, \bibinfo
		{author} {\bibfnamefont {P.~ V}\ \bibnamefont {Coveney}}, \bibinfo {author}
		{\bibfnamefont {P.~ J}\ \bibnamefont {Love}}, \bibinfo {author}
		{\bibfnamefont {H.}~\bibnamefont {Neven}}, \bibinfo {author} {\bibfnamefont
			{A.}~\bibnamefont {Aspuru-Guzik}}, \ and\ \bibinfo {author} {\bibfnamefont
			{J.~ M}\ \bibnamefont {Martinis}},\ }\bibfield  {title} {\enquote {\bibinfo
			{title} {Scalable quantum simulation of molecular energies},}\ }\href
	{\doibase 10.1103/PhysRevX.6.031007} {\bibfield  {journal} {\bibinfo
			{journal} {Physical Review X}\ }\textbf {\bibinfo {volume} {6}},\ \bibinfo
		{pages} {031007} (\bibinfo {year} {2016})}\BibitemShut {NoStop}%
	\bibitem [{\citenamefont {Li}\ and\ \citenamefont {Benjamin}(2017)}]{li_17}%
	\BibitemOpen
	\bibfield  {author} {\bibinfo {author} {\bibfnamefont {Ying}\ \bibnamefont
			{Li}}\ and\ \bibinfo {author} {\bibfnamefont {Simon~C.}\ \bibnamefont
			{Benjamin}},\ }\bibfield  {title} {\enquote {\bibinfo {title} {Efficient
				variational quantum simulator incorporating active error minimization},}\
	}\href {\doibase 10.1103/PhysRevX.7.021050} {\bibfield  {journal} {\bibinfo
		{journal} {Physical Review X}\ }\textbf {\bibinfo {volume} {7}},\ \bibinfo
	{pages} {021050} (\bibinfo {year} {2017})}\BibitemShut {NoStop}%
\bibitem [{\citenamefont {McClean}\ \emph {et~al.}(2017)\citenamefont
	{McClean}, \citenamefont {Kimchi-Schwartz}, \citenamefont {Carter},\ and\
	\citenamefont {de~Jong}}]{mcclean_17}%
\BibitemOpen
\bibfield  {author} {\bibinfo {author} {\bibfnamefont {Jarrod~R.}\
		\bibnamefont {McClean}}, \bibinfo {author} {\bibfnamefont {Mollie~E.}\
		\bibnamefont {Kimchi-Schwartz}}, \bibinfo {author} {\bibfnamefont {Jonathan}\
		\bibnamefont {Carter}}, \ and\ \bibinfo {author} {\bibfnamefont {Wibe~A.}\
		\bibnamefont {de~Jong}},\ }\bibfield  {title} {\enquote {\bibinfo {title}
		{Hybrid quantum-classical hierarchy for mitigation of decoherence and
			determination of excited states},}\ }\href {\doibase
	10.1103/PhysRevA.95.042308} {\bibfield  {journal} {\bibinfo  {journal} {Phys.
			Rev. A}\ }\textbf {\bibinfo {volume} {95}},\ \bibinfo {pages} {042308}
	(\bibinfo {year} {2017})}\BibitemShut {NoStop}%
\bibitem [{\citenamefont {Shen}\ \emph {et~al.}(2017)\citenamefont {Shen},
	\citenamefont {Zhang}, \citenamefont {Zhang}, \citenamefont {Zhang},
	\citenamefont {Yung},\ and\ \citenamefont {Kim}}]{shen_17}%
\BibitemOpen
\bibfield  {author} {\bibinfo {author} {\bibfnamefont {Yangchao}\
		\bibnamefont {Shen}}, \bibinfo {author} {\bibfnamefont {Xiang}\ \bibnamefont
		{Zhang}}, \bibinfo {author} {\bibfnamefont {Shuaining}\ \bibnamefont
		{Zhang}}, \bibinfo {author} {\bibfnamefont {Jing-Ning}\ \bibnamefont
		{Zhang}}, \bibinfo {author} {\bibfnamefont {Man-Hong}\ \bibnamefont {Yung}},
	\ and\ \bibinfo {author} {\bibfnamefont {Kihwan}\ \bibnamefont {Kim}},\
}\bibfield  {title} {\enquote {\bibinfo {title} {Quantum implementation of
		the unitary coupled cluster for simulating molecular electronic structure},}\
}\href {\doibase 10.1103/PhysRevA.95.020501} {\bibfield  {journal} {\bibinfo
	{journal} {Phys. Rev. A}\ }\textbf {\bibinfo {volume} {95}},\ \bibinfo
{pages} {020501} (\bibinfo {year} {2017})}\BibitemShut {NoStop}%
\bibitem [{\citenamefont {Kandala}\ \emph {et~al.}(2017)\citenamefont
	{Kandala}, \citenamefont {Mezzacapo}, \citenamefont {Temme}, \citenamefont
	{Takita}, \citenamefont {Brink}, \citenamefont {Chow},\ and\ \citenamefont
	{Gambetta}}]{kandala_17}%
\BibitemOpen
\bibfield  {author} {\bibinfo {author} {\bibfnamefont {Abhinav}\ \bibnamefont
		{Kandala}}, \bibinfo {author} {\bibfnamefont {Antonio}\ \bibnamefont
		{Mezzacapo}}, \bibinfo {author} {\bibfnamefont {Kristan}\ \bibnamefont
		{Temme}}, \bibinfo {author} {\bibfnamefont {Maika}\ \bibnamefont {Takita}},
	\bibinfo {author} {\bibfnamefont {Markus}\ \bibnamefont {Brink}}, \bibinfo
	{author} {\bibfnamefont {Jerry~M.}\ \bibnamefont {Chow}}, \ and\ \bibinfo
	{author} {\bibfnamefont {Jay~M.}\ \bibnamefont {Gambetta}},\ }\bibfield
{title} {\enquote {\bibinfo {title} {Hardware-efficient variational quantum
			eigensolver for small molecules and quantum magnets},}\ }\href {\doibase
	10.1038/nature23879} {\bibfield  {journal} {\bibinfo  {journal} {Nature}\
	}\textbf {\bibinfo {volume} {549}},\ \bibinfo {pages} {242--246} (\bibinfo
	{year} {2017})}\BibitemShut {NoStop}%
\bibitem [{\citenamefont {Hempel}\ \emph {et~al.}(2018)\citenamefont {Hempel},
	\citenamefont {Maier}, \citenamefont {Romero}, \citenamefont {McClean},
	\citenamefont {Monz}, \citenamefont {Shen}, \citenamefont {Jurcevic},
	\citenamefont {Lanyon}, \citenamefont {Love}, \citenamefont {Babbush},
	\citenamefont {Aspuru-Guzik}, \citenamefont {Blatt},\ and\ \citenamefont
	{Roos}}]{hempel_18}%
\BibitemOpen
\bibfield  {author} {\bibinfo {author} {\bibfnamefont {Cornelius}\
		\bibnamefont {Hempel}}, \bibinfo {author} {\bibfnamefont {Christine}\
		\bibnamefont {Maier}}, \bibinfo {author} {\bibfnamefont {Jonathan}\
		\bibnamefont {Romero}}, \bibinfo {author} {\bibfnamefont {Jarrod}\
		\bibnamefont {McClean}}, \bibinfo {author} {\bibfnamefont {Thomas}\
		\bibnamefont {Monz}}, \bibinfo {author} {\bibfnamefont {Heng}\ \bibnamefont
		{Shen}}, \bibinfo {author} {\bibfnamefont {Petar}\ \bibnamefont {Jurcevic}},
	\bibinfo {author} {\bibfnamefont {Ben~P.}\ \bibnamefont {Lanyon}}, \bibinfo
	{author} {\bibfnamefont {Peter}\ \bibnamefont {Love}}, \bibinfo {author}
	{\bibfnamefont {Ryan}\ \bibnamefont {Babbush}}, \bibinfo {author}
	{\bibfnamefont {Alán}\ \bibnamefont {Aspuru-Guzik}}, \bibinfo {author}
	{\bibfnamefont {Rainer}\ \bibnamefont {Blatt}}, \ and\ \bibinfo {author}
	{\bibfnamefont {Christian~F.}\ \bibnamefont {Roos}},\ }\bibfield  {title}
{\enquote {\bibinfo {title} {Quantum chemistry calculations on a trapped-ion
			quantum simulator},}\ }\href {\doibase 10.1103/PhysRevX.8.031022} {\bibfield
	{journal} {\bibinfo  {journal} {Physical Review X}\ }\textbf {\bibinfo
		{volume} {8}},\ \bibinfo {pages} {031022} (\bibinfo {year}
	{2018})}\BibitemShut {NoStop}%
\bibitem [{\citenamefont {Anschuetz}\ \emph {et~al.}(2018)\citenamefont
	{Anschuetz}, \citenamefont {Olson}, \citenamefont {Aspuru-Guzik},\ and\
	\citenamefont {Cao}}]{anschuetz_18}%
\BibitemOpen
\bibfield  {author} {\bibinfo {author} {\bibfnamefont {Eric~R.}\ \bibnamefont
		{Anschuetz}}, \bibinfo {author} {\bibfnamefont {Jonathan~P.}\ \bibnamefont
		{Olson}}, \bibinfo {author} {\bibfnamefont {Alán}\ \bibnamefont
		{Aspuru-Guzik}}, \ and\ \bibinfo {author} {\bibfnamefont {Yudong}\
		\bibnamefont {Cao}},\ }\bibfield  {title} {\enquote {\bibinfo {title}
		{Variational quantum factoring},}\ }\href {http://arXiv.org/abs/1808.08927}
{\bibfield  {journal} {\bibinfo  {journal} {arXiv:1808.08927}\ } (\bibinfo
	{year} {2018})}\BibitemShut {NoStop}%
\bibitem [{\citenamefont {Mitarai}\ \emph {et~al.}(2018)\citenamefont
	{Mitarai}, \citenamefont {Negoro}, \citenamefont {Kitagawa},\ and\
	\citenamefont {Fujii}}]{mitarai_18}%
\BibitemOpen
\bibfield  {author} {\bibinfo {author} {\bibfnamefont {K.}~\bibnamefont
		{Mitarai}}, \bibinfo {author} {\bibfnamefont {M.}~\bibnamefont {Negoro}},
	\bibinfo {author} {\bibfnamefont {M.}~\bibnamefont {Kitagawa}}, \ and\
	\bibinfo {author} {\bibfnamefont {K.}~\bibnamefont {Fujii}},\ }\bibfield
{title} {\enquote {\bibinfo {title} {Quantum circuit learning},}\ }\href
{\doibase 10.1103/PhysRevA.98.032309} {\bibfield  {journal} {\bibinfo
		{journal} {Phys. Rev. A}\ }\textbf {\bibinfo {volume} {98}},\ \bibinfo
	{pages} {032309} (\bibinfo {year} {2018})}\BibitemShut {NoStop}%
\bibitem [{\citenamefont {Moll}\ \emph {et~al.}(2018)\citenamefont {Moll},
	\citenamefont {Barkoutsos}, \citenamefont {Bishop}, \citenamefont {Chow},
	\citenamefont {Cross}, \citenamefont {Egger}, \citenamefont {Filipp},
	\citenamefont {Fuhrer}, \citenamefont {Gambetta}, \citenamefont {Ganzhorn},
	\citenamefont {Kandala}, \citenamefont {Mezzacapo}, \citenamefont {Müller},
	\citenamefont {Riess}, \citenamefont {Salis}, \citenamefont {Smolin},
	\citenamefont {Tavernelli},\ and\ \citenamefont {Temme}}]{moll_18}%
\BibitemOpen
\bibfield  {author} {\bibinfo {author} {\bibfnamefont {Nikolaj}\ \bibnamefont
		{Moll}}, \bibinfo {author} {\bibfnamefont {Panagiotis}\ \bibnamefont
		{Barkoutsos}}, \bibinfo {author} {\bibfnamefont {Lev~S.}\ \bibnamefont
		{Bishop}}, \bibinfo {author} {\bibfnamefont {Jerry~M.}\ \bibnamefont {Chow}},
	\bibinfo {author} {\bibfnamefont {Andrew}\ \bibnamefont {Cross}}, \bibinfo
	{author} {\bibfnamefont {Daniel~J.}\ \bibnamefont {Egger}}, \bibinfo {author}
	{\bibfnamefont {Stefan}\ \bibnamefont {Filipp}}, \bibinfo {author}
	{\bibfnamefont {Andreas}\ \bibnamefont {Fuhrer}}, \bibinfo {author}
	{\bibfnamefont {Jay~M.}\ \bibnamefont {Gambetta}}, \bibinfo {author}
	{\bibfnamefont {Marc}\ \bibnamefont {Ganzhorn}}, \bibinfo {author}
	{\bibfnamefont {Abhinav}\ \bibnamefont {Kandala}}, \bibinfo {author}
	{\bibfnamefont {Antonio}\ \bibnamefont {Mezzacapo}}, \bibinfo {author}
	{\bibfnamefont {Peter}\ \bibnamefont {Müller}}, \bibinfo {author}
	{\bibfnamefont {Walter}\ \bibnamefont {Riess}}, \bibinfo {author}
	{\bibfnamefont {Gian}\ \bibnamefont {Salis}}, \bibinfo {author}
	{\bibfnamefont {John}\ \bibnamefont {Smolin}}, \bibinfo {author}
	{\bibfnamefont {Ivano}\ \bibnamefont {Tavernelli}}, \ and\ \bibinfo {author}
	{\bibfnamefont {Kristan}\ \bibnamefont {Temme}},\ }\bibfield  {title}
{\enquote {\bibinfo {title} {Quantum optimization using variational
			algorithms on near-term quantum devices},}\ }\href {\doibase
	10.1088/2058-9565/aab822} {\bibfield  {journal} {\bibinfo  {journal} {Quantum
			Science and Technology}\ }\textbf {\bibinfo {volume} {3}},\ \bibinfo {pages}
	{030503} (\bibinfo {year} {2018})}\BibitemShut {NoStop}%
\bibitem [{\citenamefont {Endo}\ \emph {et~al.}(2018)\citenamefont {Endo},
	\citenamefont {Sun}, \citenamefont {Li}, \citenamefont {Benjamin},\ and\
	\citenamefont {Yuan}}]{endo2018variational}%
\BibitemOpen
\bibfield  {author} {\bibinfo {author} {\bibfnamefont {Suguru}\ \bibnamefont
		{Endo}}, \bibinfo {author} {\bibfnamefont {Jinzhao}\ \bibnamefont {Sun}},
	\bibinfo {author} {\bibfnamefont {Ying}\ \bibnamefont {Li}}, \bibinfo
	{author} {\bibfnamefont {Simon}\ \bibnamefont {Benjamin}}, \ and\ \bibinfo
	{author} {\bibfnamefont {Xiao}\ \bibnamefont {Yuan}},\ }\href@noop {}
{\enquote {\bibinfo {title} {Variational quantum simulation of general
			processes},}\ } (\bibinfo {year} {2018}),\ \Eprint
{http://arxiv.org/abs/1812.08778} {arXiv:1812.08778 [quant-ph]} \BibitemShut
{NoStop}%
\bibitem [{\citenamefont {Kokail}\ \emph {et~al.}(2019)\citenamefont {Kokail},
	\citenamefont {Maier}, \citenamefont {van Bijnen}, \citenamefont {Brydges},
	\citenamefont {Joshi}, \citenamefont {Jurcevic}, \citenamefont {Muschik},
	\citenamefont {Silvi}, \citenamefont {Blatt}, \citenamefont {Roos},\ and\
	\citenamefont {Zoller}}]{kokail_19}%
\BibitemOpen
\bibfield  {author} {\bibinfo {author} {\bibfnamefont {C.}~\bibnamefont
		{Kokail}}, \bibinfo {author} {\bibfnamefont {C.}~\bibnamefont {Maier}},
	\bibinfo {author} {\bibfnamefont {R.}~\bibnamefont {van Bijnen}}, \bibinfo
	{author} {\bibfnamefont {T.}~\bibnamefont {Brydges}}, \bibinfo {author}
	{\bibfnamefont {M.~K.}\ \bibnamefont {Joshi}}, \bibinfo {author}
	{\bibfnamefont {P.}~\bibnamefont {Jurcevic}}, \bibinfo {author}
	{\bibfnamefont {C.~A.}\ \bibnamefont {Muschik}}, \bibinfo {author}
	{\bibfnamefont {P.}~\bibnamefont {Silvi}}, \bibinfo {author} {\bibfnamefont
		{R.}~\bibnamefont {Blatt}}, \bibinfo {author} {\bibfnamefont {C.~F.}\
		\bibnamefont {Roos}}, \ and\ \bibinfo {author} {\bibfnamefont
		{P.}~\bibnamefont {Zoller}},\ }\bibfield  {title} {\enquote {\bibinfo {title}
		{Self-verifying variational quantum simulation of lattice models},}\ }\href
{\doibase 10.1038/s41586-019-1177-4} {\bibfield  {journal} {\bibinfo
		{journal} {Nature}\ }\textbf {\bibinfo {volume} {569}},\ \bibinfo {pages}
	{355--360} (\bibinfo {year} {2019})}\BibitemShut {NoStop}%
\bibitem [{\citenamefont {Takeshita}\ \emph {et~al.}(2020)\citenamefont
	{Takeshita}, \citenamefont {Rubin}, \citenamefont {Jiang}, \citenamefont
	{Lee}, \citenamefont {Babbush},\ and\ \citenamefont
	{McClean}}]{takeshita_19}%
\BibitemOpen
\bibfield  {author} {\bibinfo {author} {\bibfnamefont {Tyler}\ \bibnamefont
		{Takeshita}}, \bibinfo {author} {\bibfnamefont {Nicholas~C.}\ \bibnamefont
		{Rubin}}, \bibinfo {author} {\bibfnamefont {Zhang}\ \bibnamefont {Jiang}},
	\bibinfo {author} {\bibfnamefont {Eunseok}\ \bibnamefont {Lee}}, \bibinfo
	{author} {\bibfnamefont {Ryan}\ \bibnamefont {Babbush}}, \ and\ \bibinfo
	{author} {\bibfnamefont {Jarrod~R.}\ \bibnamefont {McClean}},\ }\bibfield
{title} {\enquote {\bibinfo {title} {Increasing the representation accuracy
			of quantum simulations of chemistry without extra quantum resources},}\
}\href {\doibase 10.1103/PhysRevX.10.011004} {\bibfield  {journal} {\bibinfo
	{journal} {Phys. Rev. X}\ }\textbf {\bibinfo {volume} {10}},\ \bibinfo
{pages} {011004} (\bibinfo {year} {2020})}\BibitemShut {NoStop}%
\bibitem [{\citenamefont {McArdle}\ \emph {et~al.}(2019)\citenamefont
	{McArdle}, \citenamefont {Jones}, \citenamefont {Endo}, \citenamefont {Li},
	\citenamefont {Benjamin},\ and\ \citenamefont {Yuan}}]{mcardle_19}%
\BibitemOpen
\bibfield  {author} {\bibinfo {author} {\bibfnamefont {Sam}\ \bibnamefont
		{McArdle}}, \bibinfo {author} {\bibfnamefont {Tyson}\ \bibnamefont {Jones}},
	\bibinfo {author} {\bibfnamefont {Suguru}\ \bibnamefont {Endo}}, \bibinfo
	{author} {\bibfnamefont {Ying}\ \bibnamefont {Li}}, \bibinfo {author}
	{\bibfnamefont {Simon~C.}\ \bibnamefont {Benjamin}}, \ and\ \bibinfo {author}
	{\bibfnamefont {Xiao}\ \bibnamefont {Yuan}},\ }\bibfield  {title} {\enquote
	{\bibinfo {title} {Variational ansatz-based quantum simulation of imaginary
			time evolution},}\ }\href {\doibase 10.1038/s41534-019-0187-2} {\bibfield
	{journal} {\bibinfo  {journal} {npj Quantum Information}\ }\textbf {\bibinfo
		{volume} {5}},\ \bibinfo {pages} {75} (\bibinfo {year} {2019})}\BibitemShut
{NoStop}%
\bibitem [{\citenamefont {Higgott}\ \emph {et~al.}(2019)\citenamefont
	{Higgott}, \citenamefont {Wang},\ and\ \citenamefont
	{Brierley}}]{Higgott_19}%
\BibitemOpen
\bibfield  {author} {\bibinfo {author} {\bibfnamefont {Oscar}\ \bibnamefont
		{Higgott}}, \bibinfo {author} {\bibfnamefont {Daochen}\ \bibnamefont {Wang}},
	\ and\ \bibinfo {author} {\bibfnamefont {Stephen}\ \bibnamefont {Brierley}},\
}\bibfield  {title} {\enquote {\bibinfo {title} {Variational {Q}uantum
		{C}omputation of {E}xcited {S}tates},}\ }\href {\doibase
10.22331/q-2019-07-01-156} {\bibfield  {journal} {\bibinfo  {journal}
	{{Quantum}}\ }\textbf {\bibinfo {volume} {3}},\ \bibinfo {pages} {156}
(\bibinfo {year} {2019})}\BibitemShut {NoStop}%
\bibitem [{\citenamefont {Wu}\ and\ \citenamefont {Hsieh}(2019)}]{wu_19}%
\BibitemOpen
\bibfield  {author} {\bibinfo {author} {\bibfnamefont {Jingxiang}\
		\bibnamefont {Wu}}\ and\ \bibinfo {author} {\bibfnamefont {Timothy~H.}\
		\bibnamefont {Hsieh}},\ }\bibfield  {title} {\enquote {\bibinfo {title}
		{Variational thermal quantum simulation via thermofield double states},}\
}\href {\doibase 10.1103/PhysRevLett.123.220502} {\bibfield  {journal}
{\bibinfo  {journal} {Phys. Rev. Lett.}\ }\textbf {\bibinfo {volume} {123}},\
\bibinfo {pages} {220502} (\bibinfo {year} {2019})}\BibitemShut {NoStop}%
\bibitem [{\citenamefont {Verdon}\ \emph {et~al.}(2019)\citenamefont {Verdon},
	\citenamefont {Broughton}, \citenamefont {McClean}, \citenamefont {Sung},
	\citenamefont {Babbush}, \citenamefont {Jiang}, \citenamefont {Neven},\ and\
	\citenamefont {Mohseni}}]{verdon_19}%
\BibitemOpen
\bibfield  {author} {\bibinfo {author} {\bibfnamefont {Guillaume}\
		\bibnamefont {Verdon}}, \bibinfo {author} {\bibfnamefont {Michael}\
		\bibnamefont {Broughton}}, \bibinfo {author} {\bibfnamefont {Jarrod~R.}\
		\bibnamefont {McClean}}, \bibinfo {author} {\bibfnamefont {Kevin~J.}\
		\bibnamefont {Sung}}, \bibinfo {author} {\bibfnamefont {Ryan}\ \bibnamefont
		{Babbush}}, \bibinfo {author} {\bibfnamefont {Zhang}\ \bibnamefont {Jiang}},
	\bibinfo {author} {\bibfnamefont {Hartmut}\ \bibnamefont {Neven}}, \ and\
	\bibinfo {author} {\bibfnamefont {Masoud}\ \bibnamefont {Mohseni}},\
}\bibfield  {title} {\enquote {\bibinfo {title} {Learning to learn with
		quantum neural networks via classical neural networks},}\ }\href
{http://arXiv.org/abs/1907.05415} {\bibfield  {journal} {\bibinfo  {journal}
		{arXiv:1907.05415}\ } (\bibinfo {year} {2019})}\BibitemShut {NoStop}%
\bibitem [{\citenamefont {Liu}\ \emph {et~al.}(2019{\natexlab{a}})\citenamefont
	{Liu}, \citenamefont {Zhang}, \citenamefont {Wan},\ and\ \citenamefont
	{Wang}}]{liu_19}%
\BibitemOpen
\bibfield  {author} {\bibinfo {author} {\bibfnamefont {Jin-Guo}\ \bibnamefont
		{Liu}}, \bibinfo {author} {\bibfnamefont {Yi-Hong}\ \bibnamefont {Zhang}},
	\bibinfo {author} {\bibfnamefont {Yuan}\ \bibnamefont {Wan}}, \ and\ \bibinfo
	{author} {\bibfnamefont {Lei}\ \bibnamefont {Wang}},\ }\bibfield  {title}
{\enquote {\bibinfo {title} {Variational quantum eigensolver with fewer
			qubits},}\ }\href {\doibase 10.1103/PhysRevResearch.1.023025} {\bibfield
	{journal} {\bibinfo  {journal} {Phys. Rev. Research}\ }\textbf {\bibinfo
		{volume} {1}},\ \bibinfo {pages} {023025} (\bibinfo {year}
	{2019}{\natexlab{a}})}\BibitemShut {NoStop}%
\bibitem [{\citenamefont {Liu}\ \emph {et~al.}(2019{\natexlab{b}})\citenamefont
	{Liu}, \citenamefont {Mao}, \citenamefont {Zhang},\ and\ \citenamefont
	{Wang}}]{liu_19_gibbs}%
\BibitemOpen
\bibfield  {author} {\bibinfo {author} {\bibfnamefont {Jin-Guo}\ \bibnamefont
		{Liu}}, \bibinfo {author} {\bibfnamefont {Liang}\ \bibnamefont {Mao}},
	\bibinfo {author} {\bibfnamefont {Pan}\ \bibnamefont {Zhang}}, \ and\
	\bibinfo {author} {\bibfnamefont {Lei}\ \bibnamefont {Wang}},\ }\href@noop {}
{\enquote {\bibinfo {title} {Solving quantum statistical mechanics with
			variational autoregressive networks and quantum circuits},}\ } (\bibinfo
{year} {2019}{\natexlab{b}}),\ \Eprint {http://arxiv.org/abs/1912.11381}
{arXiv:1912.11381 [quant-ph]} \BibitemShut {NoStop}%
\bibitem [{\citenamefont {Benedetti}\ \emph {et~al.}(2019)\citenamefont
	{Benedetti}, \citenamefont {Lloyd}, \citenamefont {Sack},\ and\ \citenamefont
	{Fiorentini}}]{benedetti_19}%
\BibitemOpen
\bibfield  {author} {\bibinfo {author} {\bibfnamefont {Marcello}\
		\bibnamefont {Benedetti}}, \bibinfo {author} {\bibfnamefont {Erika}\
		\bibnamefont {Lloyd}}, \bibinfo {author} {\bibfnamefont {Stefan}\
		\bibnamefont {Sack}}, \ and\ \bibinfo {author} {\bibfnamefont {Mattia}\
		\bibnamefont {Fiorentini}},\ }\bibfield  {title} {\enquote {\bibinfo {title}
		{Parameterized quantum circuits as machine learning models},}\ }\href
{\doibase 10.1088/2058-9565/ab4eb5} {\bibfield  {journal} {\bibinfo
		{journal} {Quantum Science and Technology}\ }\textbf {\bibinfo {volume}
		{4}},\ \bibinfo {pages} {043001} (\bibinfo {year} {2019})}\BibitemShut
{NoStop}%
\bibitem [{\citenamefont {Grimsley}\ \emph {et~al.}(2019)\citenamefont
	{Grimsley}, \citenamefont {Economou}, \citenamefont {Barnes},\ and\
	\citenamefont {Mayhall}}]{grimsley_19}%
\BibitemOpen
\bibfield  {author} {\bibinfo {author} {\bibfnamefont {Harper~R.}\
		\bibnamefont {Grimsley}}, \bibinfo {author} {\bibfnamefont {Sophia~E.}\
		\bibnamefont {Economou}}, \bibinfo {author} {\bibfnamefont {Edwin}\
		\bibnamefont {Barnes}}, \ and\ \bibinfo {author} {\bibfnamefont
		{Nicholas~J.}\ \bibnamefont {Mayhall}},\ }\bibfield  {title} {\enquote
	{\bibinfo {title} {An adaptive variational algorithm for exact molecular
			simulations on a quantum computer},}\ }\href {\doibase
	10.1038/s41467-019-10988-2} {\bibfield  {journal} {\bibinfo  {journal}
		{Nature Communications}\ }\textbf {\bibinfo {volume} {10}},\ \bibinfo {pages}
	{3007} (\bibinfo {year} {2019})}\BibitemShut {NoStop}%
\bibitem [{\citenamefont {Chowdhury}\ \emph {et~al.}(2020)\citenamefont
	{Chowdhury}, \citenamefont {Low},\ and\ \citenamefont
	{Wiebe}}]{chowdhury2020variational}%
\BibitemOpen
\bibfield  {author} {\bibinfo {author} {\bibfnamefont {Anirban~N.}\
		\bibnamefont {Chowdhury}}, \bibinfo {author} {\bibfnamefont {Guang~Hao}\
		\bibnamefont {Low}}, \ and\ \bibinfo {author} {\bibfnamefont {Nathan}\
		\bibnamefont {Wiebe}},\ }\href@noop {} {\enquote {\bibinfo {title} {A
			variational quantum algorithm for preparing quantum gibbs states},}\ }
(\bibinfo {year} {2020}),\ \Eprint {http://arxiv.org/abs/2002.00055}
{arXiv:2002.00055 [quant-ph]} \BibitemShut {NoStop}%
\bibitem [{\citenamefont {Wang}\ \emph {et~al.}(2020)\citenamefont {Wang},
	\citenamefont {Li},\ and\ \citenamefont {Wang}}]{wang_20}%
\BibitemOpen
\bibfield  {author} {\bibinfo {author} {\bibfnamefont {Youle}\ \bibnamefont
		{Wang}}, \bibinfo {author} {\bibfnamefont {Guangxi}\ \bibnamefont {Li}}, \
	and\ \bibinfo {author} {\bibfnamefont {Xin}\ \bibnamefont {Wang}},\
}\href@noop {} {\enquote {\bibinfo {title} {Variational quantum gibbs state
		preparation with a truncated taylor series},}\ } (\bibinfo {year} {2020}),\
\Eprint {http://arxiv.org/abs/2005.08797} {arXiv:2005.08797 [quant-ph]}
\BibitemShut {NoStop}%
\bibitem [{\citenamefont {Zhang}\ and\ \citenamefont {Yin}(2020)}]{danbo_20}%
\BibitemOpen
\bibfield  {author} {\bibinfo {author} {\bibfnamefont {Dan-Bo}\ \bibnamefont
		{Zhang}}\ and\ \bibinfo {author} {\bibfnamefont {Tao}\ \bibnamefont {Yin}},\
}\bibfield  {title} {\enquote {\bibinfo {title} {Collective optimization for
		variational quantum eigensolvers},}\ }\href {\doibase
10.1103/PhysRevA.101.032311} {\bibfield  {journal} {\bibinfo  {journal}
	{Phys. Rev. A}\ }\textbf {\bibinfo {volume} {101}},\ \bibinfo {pages}
{032311} (\bibinfo {year} {2020})}\BibitemShut {NoStop}%
\bibitem [{\citenamefont {Arute}\ \emph {et~al.}(2020)\citenamefont {Arute}
	\emph {et~al.}}]{arute2020hartreefock}%
\BibitemOpen
\bibfield  {author} {\bibinfo {author} {\bibfnamefont {Frank}\ \bibnamefont
		{Arute}} \emph {et~al.},\ }\href@noop {} {\enquote {\bibinfo {title}
		{Hartree-fock on a superconducting qubit quantum computer},}\ } (\bibinfo
{year} {2020}),\ \Eprint {http://arxiv.org/abs/2004.04174} {arXiv:2004.04174
	[quant-ph]} \BibitemShut {NoStop}%
\bibitem [{\citenamefont {Dallaire-Demers}\ \emph {et~al.}(2020)\citenamefont
	{Dallaire-Demers}, \citenamefont {Stęchły}, \citenamefont {Gonthier},
	\citenamefont {Bashige}, \citenamefont {Romero},\ and\ \citenamefont
	{Cao}}]{dallairedemers2020application}%
\BibitemOpen
\bibfield  {author} {\bibinfo {author} {\bibfnamefont {Pierre-Luc}\
		\bibnamefont {Dallaire-Demers}}, \bibinfo {author} {\bibfnamefont {Michał}\
		\bibnamefont {Stęchły}}, \bibinfo {author} {\bibfnamefont {Jerome~F.}\
		\bibnamefont {Gonthier}}, \bibinfo {author} {\bibfnamefont
		{Ntwali~Toussaint}\ \bibnamefont {Bashige}}, \bibinfo {author} {\bibfnamefont
		{Jonathan}\ \bibnamefont {Romero}}, \ and\ \bibinfo {author} {\bibfnamefont
		{Yudong}\ \bibnamefont {Cao}},\ }\href@noop {} {\enquote {\bibinfo {title}
		{An application benchmark for fermionic quantum simulations},}\ } (\bibinfo
{year} {2020}),\ \Eprint {http://arxiv.org/abs/2003.01862} {arXiv:2003.01862
	[quant-ph]} \BibitemShut {NoStop}%
\bibitem [{\citenamefont {Xu}\ \emph {et~al.}(2019)\citenamefont {Xu},
	\citenamefont {Sun}, \citenamefont {Endo}, \citenamefont {Li}, \citenamefont
	{Benjamin},\ and\ \citenamefont {Yuan}}]{xu2019variational}%
\BibitemOpen
\bibfield  {author} {\bibinfo {author} {\bibfnamefont {Xiaosi}\ \bibnamefont
		{Xu}}, \bibinfo {author} {\bibfnamefont {Jinzhao}\ \bibnamefont {Sun}},
	\bibinfo {author} {\bibfnamefont {Suguru}\ \bibnamefont {Endo}}, \bibinfo
	{author} {\bibfnamefont {Ying}\ \bibnamefont {Li}}, \bibinfo {author}
	{\bibfnamefont {Simon~C.}\ \bibnamefont {Benjamin}}, \ and\ \bibinfo {author}
	{\bibfnamefont {Xiao}\ \bibnamefont {Yuan}},\ }\href@noop {} {\enquote
	{\bibinfo {title} {Variational algorithms for linear algebra},}\ } (\bibinfo
{year} {2019}),\ \Eprint {http://arxiv.org/abs/1909.03898} {arXiv:1909.03898
	[quant-ph]} \BibitemShut {NoStop}%
\bibitem [{\citenamefont {Lubasch}\ \emph {et~al.}(2020)\citenamefont
	{Lubasch}, \citenamefont {Joo}, \citenamefont {Moinier}, \citenamefont
	{Kiffner},\ and\ \citenamefont {Jaksch}}]{lubasch_20}%
\BibitemOpen
\bibfield  {author} {\bibinfo {author} {\bibfnamefont {Michael}\ \bibnamefont
		{Lubasch}}, \bibinfo {author} {\bibfnamefont {Jaewoo}\ \bibnamefont {Joo}},
	\bibinfo {author} {\bibfnamefont {Pierre}\ \bibnamefont {Moinier}}, \bibinfo
	{author} {\bibfnamefont {Martin}\ \bibnamefont {Kiffner}}, \ and\ \bibinfo
	{author} {\bibfnamefont {Dieter}\ \bibnamefont {Jaksch}},\ }\bibfield
{title} {\enquote {\bibinfo {title} {Variational quantum algorithms for
			nonlinear problems},}\ }\href {\doibase 10.1103/PhysRevA.101.010301}
{\bibfield  {journal} {\bibinfo  {journal} {Phys. Rev. A}\ }\textbf {\bibinfo
		{volume} {101}},\ \bibinfo {pages} {010301} (\bibinfo {year}
	{2020})}\BibitemShut {NoStop}%
\bibitem [{\citenamefont {Greene-Diniz}\ and\ \citenamefont
	{Ramo}(2019)}]{greenediniz2019generalized}%
\BibitemOpen
\bibfield  {author} {\bibinfo {author} {\bibfnamefont {Gabriel}\ \bibnamefont
		{Greene-Diniz}}\ and\ \bibinfo {author} {\bibfnamefont {David~Muñoz}\
		\bibnamefont {Ramo}},\ }\href@noop {} {\enquote {\bibinfo {title}
		{Generalized unitary coupled cluster excitations for multireference molecular
			states optimized by the variational quantum eigensolver},}\ } (\bibinfo
{year} {2019}),\ \Eprint {http://arxiv.org/abs/1910.05168} {arXiv:1910.05168
	[quant-ph]} \BibitemShut {NoStop}%
\bibitem [{\citenamefont {Nakanishi}\ \emph {et~al.}(2019)\citenamefont
	{Nakanishi}, \citenamefont {Mitarai},\ and\ \citenamefont
	{Fujii}}]{nakanishi_19}%
\BibitemOpen
\bibfield  {author} {\bibinfo {author} {\bibfnamefont {Ken~M.}\ \bibnamefont
		{Nakanishi}}, \bibinfo {author} {\bibfnamefont {Kosuke}\ \bibnamefont
		{Mitarai}}, \ and\ \bibinfo {author} {\bibfnamefont {Keisuke}\ \bibnamefont
		{Fujii}},\ }\bibfield  {title} {\enquote {\bibinfo {title} {Subspace-search
			variational quantum eigensolver for excited states},}\ }\href {\doibase
	10.1103/PhysRevResearch.1.033062} {\bibfield  {journal} {\bibinfo  {journal}
		{Physical Review Research}\ }\textbf {\bibinfo {volume} {1}},\ \bibinfo
	{pages} {033062} (\bibinfo {year} {2019})}\BibitemShut {NoStop}%
\bibitem [{\citenamefont {Bartlett}\ \emph {et~al.}(1935)\citenamefont
	{Bartlett}, \citenamefont {Gibbons},\ and\ \citenamefont
	{Dunn}}]{bartlett_35}%
\BibitemOpen
\bibfield  {author} {\bibinfo {author} {\bibfnamefont {J.~H.}\ \bibnamefont
		{Bartlett}}, \bibinfo {author} {\bibfnamefont {J.~J.}\ \bibnamefont
		{Gibbons}}, \ and\ \bibinfo {author} {\bibfnamefont {C.~G.}\ \bibnamefont
		{Dunn}},\ }\bibfield  {title} {\enquote {\bibinfo {title} {The normal helium
			atom},}\ }\href {\doibase 10.1103/PhysRev.47.679} {\bibfield  {journal}
	{\bibinfo  {journal} {Physical Review}\ }\textbf {\bibinfo {volume} {47}},\
	\bibinfo {pages} {679--680} (\bibinfo {year} {1935})}\BibitemShut {NoStop}%
\bibitem [{\citenamefont {Siringo}\ and\ \citenamefont
	{Marotta}(2005)}]{siringo_05}%
\BibitemOpen
\bibfield  {author} {\bibinfo {author} {\bibfnamefont {F.}~\bibnamefont
		{Siringo}}\ and\ \bibinfo {author} {\bibfnamefont {L.}~\bibnamefont
		{Marotta}},\ }\bibfield  {title} {\enquote {\bibinfo {title} {A variational
			method from the variance of energy},}\ }\href {\doibase
	10.1140/epjc/s2005-02358-x} {\bibfield  {journal} {\bibinfo  {journal} {The
			European Physical Journal C - Particles and Fields}\ }\textbf {\bibinfo
		{volume} {44}},\ \bibinfo {pages} {293--298} (\bibinfo {year}
	{2005})}\BibitemShut {NoStop}%
\bibitem [{\citenamefont {Umrigar}\ \emph {et~al.}(1988)\citenamefont
	{Umrigar}, \citenamefont {Wilson},\ and\ \citenamefont
	{Wilkins}}]{umrigar_88}%
\BibitemOpen
\bibfield  {author} {\bibinfo {author} {\bibfnamefont {C.~J.}\ \bibnamefont
		{Umrigar}}, \bibinfo {author} {\bibfnamefont {K.~G.}\ \bibnamefont {Wilson}},
	\ and\ \bibinfo {author} {\bibfnamefont {J.~W.}\ \bibnamefont {Wilkins}},\
}\bibfield  {title} {\enquote {\bibinfo {title} {Optimized trial wave
		functions for quantum monte carlo calculations},}\ }\href {\doibase
10.1103/PhysRevLett.60.1719} {\bibfield  {journal} {\bibinfo  {journal}
	{Phys. Rev. Lett.}\ }\textbf {\bibinfo {volume} {60}},\ \bibinfo {pages}
{1719--1722} (\bibinfo {year} {1988})}\BibitemShut {NoStop}%
\bibitem [{\citenamefont {Umrigar}\ and\ \citenamefont
	{Filippi}(2005)}]{umrigar_05}%
\BibitemOpen
\bibfield  {author} {\bibinfo {author} {\bibfnamefont {C.~J.}\ \bibnamefont
		{Umrigar}}\ and\ \bibinfo {author} {\bibfnamefont {Claudia}\ \bibnamefont
		{Filippi}},\ }\bibfield  {title} {\enquote {\bibinfo {title} {Energy and
			variance optimization of many-body wave functions},}\ }\href {\doibase
	10.1103/PhysRevLett.94.150201} {\bibfield  {journal} {\bibinfo  {journal}
		{Phys. Rev. Lett.}\ }\textbf {\bibinfo {volume} {94}},\ \bibinfo {pages}
	{150201} (\bibinfo {year} {2005})}\BibitemShut {NoStop}%
\bibitem [{\citenamefont {Khemani}\ \emph {et~al.}(2016)\citenamefont
	{Khemani}, \citenamefont {Pollmann},\ and\ \citenamefont
	{Sondhi}}]{khemani_16}%
\BibitemOpen
\bibfield  {author} {\bibinfo {author} {\bibfnamefont {Vedika}\ \bibnamefont
		{Khemani}}, \bibinfo {author} {\bibfnamefont {Frank}\ \bibnamefont
		{Pollmann}}, \ and\ \bibinfo {author} {\bibfnamefont {S.~ L}\ \bibnamefont
		{Sondhi}},\ }\bibfield  {title} {\enquote {\bibinfo {title} {Obtaining highly
			excited eigenstates of many-body localized hamiltonians by the density matrix
			renormalization group approach},}\ }\href {\doibase
	10.1103/PhysRevLett.116.247204} {\bibfield  {journal} {\bibinfo  {journal}
		{Phys. Rev. Lett.}\ }\textbf {\bibinfo {volume} {116}},\ \bibinfo {pages}
	{247204} (\bibinfo {year} {2016})}\BibitemShut {NoStop}%
\bibitem [{\citenamefont {Pollmann}\ \emph {et~al.}(2016)\citenamefont
	{Pollmann}, \citenamefont {Khemani}, \citenamefont {Cirac},\ and\
	\citenamefont {Sondhi}}]{pollmann_16}%
\BibitemOpen
\bibfield  {author} {\bibinfo {author} {\bibfnamefont {Frank}\ \bibnamefont
		{Pollmann}}, \bibinfo {author} {\bibfnamefont {Vedika}\ \bibnamefont
		{Khemani}}, \bibinfo {author} {\bibfnamefont {J.~Ignacio}\ \bibnamefont
		{Cirac}}, \ and\ \bibinfo {author} {\bibfnamefont {S.~L.}\ \bibnamefont
		{Sondhi}},\ }\bibfield  {title} {\enquote {\bibinfo {title} {Efficient
			variational diagonalization of fully many-body localized hamiltonians},}\
}\href {\doibase 10.1103/PhysRevB.94.041116} {\bibfield  {journal} {\bibinfo
	{journal} {Phys. Rev. B}\ }\textbf {\bibinfo {volume} {94}},\ \bibinfo
{pages} {041116} (\bibinfo {year} {2016})}\BibitemShut {NoStop}%
\bibitem [{\citenamefont {Vicentini}\ \emph {et~al.}(2019)\citenamefont
	{Vicentini}, \citenamefont {Biella}, \citenamefont {Regnault},\ and\
	\citenamefont {Ciuti}}]{vicentini_19}%
\BibitemOpen
\bibfield  {author} {\bibinfo {author} {\bibfnamefont {Filippo}\ \bibnamefont
		{Vicentini}}, \bibinfo {author} {\bibfnamefont {Alberto}\ \bibnamefont
		{Biella}}, \bibinfo {author} {\bibfnamefont {Nicolas}\ \bibnamefont
		{Regnault}}, \ and\ \bibinfo {author} {\bibfnamefont {Cristiano}\
		\bibnamefont {Ciuti}},\ }\bibfield  {title} {\enquote {\bibinfo {title}
		{Variational neural-network ansatz for steady states in open quantum
			systems},}\ }\href {\doibase 10.1103/PhysRevLett.122.250503} {\bibfield
	{journal} {\bibinfo  {journal} {Phys. Rev. Lett.}\ }\textbf {\bibinfo
		{volume} {122}},\ \bibinfo {pages} {250503} (\bibinfo {year}
	{2019})}\BibitemShut {NoStop}%
\bibitem [{\citenamefont {Bairey}\ \emph {et~al.}(2019)\citenamefont {Bairey},
	\citenamefont {Arad},\ and\ \citenamefont {Lindner}}]{bairey_19}%
\BibitemOpen
\bibfield  {author} {\bibinfo {author} {\bibfnamefont {Eyal}\ \bibnamefont
		{Bairey}}, \bibinfo {author} {\bibfnamefont {Itai}\ \bibnamefont {Arad}}, \
	and\ \bibinfo {author} {\bibfnamefont {Netanel~H.}\ \bibnamefont {Lindner}},\
}\bibfield  {title} {\enquote {\bibinfo {title} {Learning a local hamiltonian
		from local measurements},}\ }\href {\doibase 10.1103/PhysRevLett.122.020504}
{\bibfield  {journal} {\bibinfo  {journal} {Phys. Rev. Lett.}\ }\textbf
	{\bibinfo {volume} {122}},\ \bibinfo {pages} {020504} (\bibinfo {year}
	{2019})}\BibitemShut {NoStop}%
\bibitem [{\citenamefont {Chertkov}\ and\ \citenamefont
	{Clark}(2018)}]{chertkov_18}%
\BibitemOpen
\bibfield  {author} {\bibinfo {author} {\bibfnamefont {Eli}\ \bibnamefont
		{Chertkov}}\ and\ \bibinfo {author} {\bibfnamefont {Bryan~K.}\ \bibnamefont
		{Clark}},\ }\bibfield  {title} {\enquote {\bibinfo {title} {Computational
			inverse method for constructing spaces of quantum models from wave
			functions},}\ }\href {\doibase 10.1103/PhysRevX.8.031029} {\bibfield
	{journal} {\bibinfo  {journal} {Physical Review X}\ }\textbf {\bibinfo
		{volume} {8}},\ \bibinfo {pages} {031029} (\bibinfo {year}
	{2018})}\BibitemShut {NoStop}%
\bibitem [{\citenamefont {Qi}\ and\ \citenamefont
	{Ranard}(2019)}]{Qi2019determininglocal}%
\BibitemOpen
\bibfield  {author} {\bibinfo {author} {\bibfnamefont {Xiao-Liang}\
		\bibnamefont {Qi}}\ and\ \bibinfo {author} {\bibfnamefont {Daniel}\
		\bibnamefont {Ranard}},\ }\bibfield  {title} {\enquote {\bibinfo {title}
		{Determining a local {H}amiltonian from a single eigenstate},}\ }\href
{\doibase 10.22331/q-2019-07-08-159} {\bibfield  {journal} {\bibinfo
		{journal} {{Quantum}}\ }\textbf {\bibinfo {volume} {3}},\ \bibinfo {pages}
	{159} (\bibinfo {year} {2019})}\BibitemShut {NoStop}%
\bibitem [{\citenamefont {Schuld}\ \emph {et~al.}(2019)\citenamefont {Schuld},
	\citenamefont {Bergholm}, \citenamefont {Gogolin}, \citenamefont {Izaac},\
	and\ \citenamefont {Killoran}}]{schuld_19}%
\BibitemOpen
\bibfield  {author} {\bibinfo {author} {\bibfnamefont {Maria}\ \bibnamefont
		{Schuld}}, \bibinfo {author} {\bibfnamefont {Ville}\ \bibnamefont
		{Bergholm}}, \bibinfo {author} {\bibfnamefont {Christian}\ \bibnamefont
		{Gogolin}}, \bibinfo {author} {\bibfnamefont {Josh}\ \bibnamefont {Izaac}}, \
	and\ \bibinfo {author} {\bibfnamefont {Nathan}\ \bibnamefont {Killoran}},\
}\bibfield  {title} {\enquote {\bibinfo {title} {Evaluating analytic
		gradients on quantum hardware},}\ }\href {\doibase
10.1103/PhysRevA.99.032331} {\bibfield  {journal} {\bibinfo  {journal} {Phys.
		Rev. A}\ }\textbf {\bibinfo {volume} {99}},\ \bibinfo {pages} {032331}
(\bibinfo {year} {2019})}\BibitemShut {NoStop}%
\bibitem [{\citenamefont {Sweke}\ \emph {et~al.}(2019)\citenamefont {Sweke},
	\citenamefont {Wilde}, \citenamefont {Meyer}, \citenamefont {Schuld},
	\citenamefont {Fährmann}, \citenamefont {Meynard-Piganeau},\ and\
	\citenamefont {Eisert}}]{sweke_19}%
\BibitemOpen
\bibfield  {author} {\bibinfo {author} {\bibfnamefont {Ryan}\ \bibnamefont
		{Sweke}}, \bibinfo {author} {\bibfnamefont {Frederik}\ \bibnamefont {Wilde}},
	\bibinfo {author} {\bibfnamefont {Johannes}\ \bibnamefont {Meyer}}, \bibinfo
	{author} {\bibfnamefont {Maria}\ \bibnamefont {Schuld}}, \bibinfo {author}
	{\bibfnamefont {Paul~K.}\ \bibnamefont {Fährmann}}, \bibinfo {author}
	{\bibfnamefont {Barthélémy}\ \bibnamefont {Meynard-Piganeau}}, \ and\
	\bibinfo {author} {\bibfnamefont {Jens}\ \bibnamefont {Eisert}},\ }\bibfield
{title} {\enquote {\bibinfo {title} {Stochastic gradient descent for hybrid
			quantum-classical optimization},}\ }\href {http://arXiv.org/abs/1910.01155v1}
{\bibfield  {journal} {\bibinfo  {journal} {arXiv:1910.01155v1}\ } (\bibinfo
	{year} {2019})}\BibitemShut {NoStop}%
\bibitem [{\citenamefont {Huggins}\ \emph {et~al.}(2019)\citenamefont
	{Huggins}, \citenamefont {McClean}, \citenamefont {Rubin}, \citenamefont
	{Jiang}, \citenamefont {Wiebe}, \citenamefont {Whaley},\ and\ \citenamefont
	{Babbush}}]{huggins2019efficient}%
\BibitemOpen
\bibfield  {author} {\bibinfo {author} {\bibfnamefont {William~J.}\
		\bibnamefont {Huggins}}, \bibinfo {author} {\bibfnamefont {Jarrod}\
		\bibnamefont {McClean}}, \bibinfo {author} {\bibfnamefont {Nicholas}\
		\bibnamefont {Rubin}}, \bibinfo {author} {\bibfnamefont {Zhang}\ \bibnamefont
		{Jiang}}, \bibinfo {author} {\bibfnamefont {Nathan}\ \bibnamefont {Wiebe}},
	\bibinfo {author} {\bibfnamefont {K.~Birgitta}\ \bibnamefont {Whaley}}, \
	and\ \bibinfo {author} {\bibfnamefont {Ryan}\ \bibnamefont {Babbush}},\
}\href@noop {} {\enquote {\bibinfo {title} {Efficient and noise resilient
		measurements for quantum chemistry on near-term quantum computers},}\ }
(\bibinfo {year} {2019}),\ \Eprint {http://arxiv.org/abs/1907.13117}
{arXiv:1907.13117 [quant-ph]} \BibitemShut {NoStop}%
\bibitem [{\citenamefont {Cotler}\ and\ \citenamefont
	{Wilczek}(2020)}]{cotler_20}%
\BibitemOpen
\bibfield  {author} {\bibinfo {author} {\bibfnamefont {Jordan}\ \bibnamefont
		{Cotler}}\ and\ \bibinfo {author} {\bibfnamefont {Frank}\ \bibnamefont
		{Wilczek}},\ }\bibfield  {title} {\enquote {\bibinfo {title} {Quantum
			overlapping tomography},}\ }\href {\doibase 10.1103/PhysRevLett.124.100401}
{\bibfield  {journal} {\bibinfo  {journal} {Phys. Rev. Lett.}\ }\textbf
	{\bibinfo {volume} {124}},\ \bibinfo {pages} {100401} (\bibinfo {year}
	{2020})}\BibitemShut {NoStop}%
\bibitem [{\citenamefont {McClean}\ \emph {et~al.}(2020)\citenamefont {McClean}
	\emph {et~al.}}]{OpenFermion}%
\BibitemOpen
\bibfield  {author} {\bibinfo {author} {\bibfnamefont {Jarrod~R.}\
		\bibnamefont {McClean}} \emph {et~al.},\ }\bibfield  {title} {\enquote
	{\bibinfo {title} {Openfermion: the electronic structure package for quantum
			computers},}\ }\href {\doibase 10.1088/2058-9565/ab8ebc} {\bibfield
	{journal} {\bibinfo  {journal} {Quantum Science and Technology}\ }\textbf
	{\bibinfo {volume} {5}},\ \bibinfo {pages} {034014} (\bibinfo {year}
	{2020})}\BibitemShut {NoStop}%
\bibitem [{HiQ()}]{HiQ}%
\BibitemOpen
\href@noop {} {}\bibinfo {note} {Huawei HiQ team, 'Huawei HiQ: A
	High-performance Quantum Computing Simulator and Programming
	Framework',http://hiq.huaweicloud.com.}\BibitemShut {Stop}%
\bibitem [{\citenamefont {Evgeniou}\ and\ \citenamefont
	{Pontil}(2004)}]{multi_task_learning}%
\BibitemOpen
\bibfield  {author} {\bibinfo {author} {\bibfnamefont {Theodoros}\
		\bibnamefont {Evgeniou}}\ and\ \bibinfo {author} {\bibfnamefont
		{Massimiliano}\ \bibnamefont {Pontil}},\ }\bibfield  {title} {\enquote
	{\bibinfo {title} {Regularized multi--task learning},}\ }in\ \href {\doibase
	10.1145/1014052.1014067} {\emph {\bibinfo {booktitle} {Proceedings of the
			Tenth ACM SIGKDD International Conference on Knowledge Discovery and Data
			Mining}}},\ \bibinfo {series and number} {KDD ’04}\ (\bibinfo  {publisher}
{Association for Computing Machinery},\ \bibinfo {address} {New York, NY,
	USA},\ \bibinfo {year} {2004})\ p.\ \bibinfo {pages} {109–117}\BibitemShut
{NoStop}%
\bibitem [{\citenamefont {Goodfellow}\ \emph {et~al.}(2016)\citenamefont
	{Goodfellow}, \citenamefont {Bengio},\ and\ \citenamefont
	{Courville}}]{Goodfellow-et-al-2016}%
\BibitemOpen
\bibfield  {author} {\bibinfo {author} {\bibfnamefont {Ian}\ \bibnamefont
		{Goodfellow}}, \bibinfo {author} {\bibfnamefont {Yoshua}\ \bibnamefont
		{Bengio}}, \ and\ \bibinfo {author} {\bibfnamefont {Aaron}\ \bibnamefont
		{Courville}},\ }\href@noop {} {\emph {\bibinfo {title} {Deep Learning}}}\
(\bibinfo  {publisher} {MIT Press},\ \bibinfo {year} {2016})\ \bibinfo {note}
{\url{http://www.deeplearningbook.org}}\BibitemShut {NoStop}%
\bibitem [{\citenamefont {K{\"{u}}bler}\ \emph {et~al.}(2020)\citenamefont
	{K{\"{u}}bler}, \citenamefont {Arrasmith}, \citenamefont {Cincio},\ and\
	\citenamefont {Coles}}]{Kubler2020adaptiveoptimizer}%
\BibitemOpen
\bibfield  {author} {\bibinfo {author} {\bibfnamefont {Jonas~M.}\
		\bibnamefont {K{\"{u}}bler}}, \bibinfo {author} {\bibfnamefont {Andrew}\
		\bibnamefont {Arrasmith}}, \bibinfo {author} {\bibfnamefont {Lukasz}\
		\bibnamefont {Cincio}}, \ and\ \bibinfo {author} {\bibfnamefont {Patrick~J.}\
		\bibnamefont {Coles}},\ }\bibfield  {title} {\enquote {\bibinfo {title} {An
			{A}daptive {O}ptimizer for {M}easurement-{F}rugal {V}ariational
			{A}lgorithms},}\ }\href {\doibase 10.22331/q-2020-05-11-263} {\bibfield
	{journal} {\bibinfo  {journal} {{Quantum}}\ }\textbf {\bibinfo {volume}
		{4}},\ \bibinfo {pages} {263} (\bibinfo {year} {2020})}\BibitemShut {NoStop}%
\end{thebibliography}
\end{document}